\documentclass[journal]{IEEEtran}

\ifCLASSINFOpdf
\usepackage[pdftex]{graphicx}
\usepackage{graphicx}

\usepackage{flushend}

\usepackage{subfigure}

\usepackage{lmodern}
\usepackage[T1]{fontenc}

\usepackage{latexsym}
\usepackage{amssymb}
\usepackage{amsmath} 
\usepackage{enumerate}
\usepackage{mathtools}
\usepackage{hyperref}
\usepackage{multirow}
\usepackage{makecell}
\usepackage{longtable} 
\usepackage{pdflscape,array,booktabs}
\usepackage[table]{xcolor}
\usepackage{footnote}
\usepackage{hhline}
\usepackage{float}
\usepackage{textcomp}

\usepackage{cite}

\usepackage[htt]{hyphenat}

\usepackage{enumitem}
\setlist[itemize]{noitemsep, topsep=0pt, leftmargin=*}

\usepackage[font=small,skip=2pt]{caption}

\usepackage[ruled, vlined, linesnumbered]{algorithm2e}
\usepackage{tabularx}

\usepackage{graphicx}

\usepackage{listings}

\definecolor{darkcerulean}{rgb}{0.03, 0.27, 0.49}
\definecolor{darkcandyapplered}{rgb}{0.64, 0.0, 0.0}
\definecolor{darkspringgreen}{rgb}{0.09, 0.45, 0.27}
\definecolor{lightgray}{rgb}{0.83, 0.83, 0.83}

\usepackage{xcolor}

\usepackage[switch,pagewise]{lineno}

\definecolor{brightmaroon}{rgb}{0.76, 0.13, 0.28}

\definecolor{algCommColor}{rgb}{0.03, 0.27, 0.49}

\ifCLASSINFOpdf
\else
\fi

\usepackage{color,soul}

\begin{document}

%
\title{\textbf{\huge Profiling and Improving the Duty-Cycling Performance of Linux-based IoT Devices}}
%
%
%

\author{\IEEEauthorblockN{Immanuel Amirtharaj,
Tai Groot, and Behnam Dezfouli\IEEEauthorrefmark{1}\thanks{\IEEEauthorrefmark{1}Corresponding Author}}\\
\IEEEauthorblockA{Internet of Things Research Lab, Department of Computer Engineering, Santa Clara University, USA
\\
iamirtharaj@scu.edu,
agroot@scu.edu,
bdezfouli@scu.edu
}}

\maketitle

\begin{abstract}

Minimizing the energy consumption of Linux-based devices is an essential step towards their wide deployment in various IoT scenarios.
Energy saving methods such as duty-cycling aim to address this constraint by limiting the amount of time the device is powered on.
In this work we study and improve the amount of time a Linux-based IoT device is powered on to accomplish its tasks.
We analyze the processes of system boot up and shutdown on two platforms, the Raspberry Pi 3 and Raspberry Pi Zero Wireless, and enhance duty-cycling performance by identifying and disabling time-consuming or unnecessary units initialized in the userspace.
We also study whether SD card speed and SD card capacity utilization affect boot up duration and energy consumption.
In addition, we propose Pallex, a parallel execution framework built on top of the \texttt{systemd init} system to run a user application concurrently with userspace initialization.
We validate the performance impact of Pallex when applied to various IoT application scenarios: (i) capturing an image, (ii) capturing and encrypting an image, (iii) capturing and classifying an image using the the k-nearest neighbor algorithm, and (iv) capturing images and sending them to a cloud server.
Our results show that system lifetime is increased by 18.3\%, 16.8\%, 13.9\% and 30.2\%, for these application scenarios, respectively.
\end{abstract}


%
\IEEEpeerreviewmaketitle

\section{Introduction}


%
%
%
%

As more Internet of Things (IoT) devices are deployed every day, the demand for energy efficient and low-power devices is increasing at a fast pace. 
These devices are increasingly deployed in remote regions or harsh environments where the only reliable source of power is a battery or the power provided by an energy harvesting solution. 
In addition, even for those devices that are connected to the power grid, it is essential to reduce their energy footprint given the increasing cost of energy and the efforts towards reducing carbon emissions \cite{delforge2016slashing,chu2012opportunities}.
To this end, various solutions have been proposed, such as energy-efficient hardware design \cite{wang2015energy,ueki2015low,CYW43907}, low-overhead operating systems (OS) \cite{baccelli2013riot, levis2005tinyos, ThreadX, FreeRTOS}, and low-power networking stacks \cite{jones2001survey,dezfouli2017rewimo,dezfouli2015dicsa,zoican2012lwip}.
The importance of energy-efficient design has also been demonstrated by mining the questions raised by developers in forums \cite{pinto2014mining}.

With the increases in processing power and memory capacity for embedded systems, new process-intensive IoT applications are being introduced.
Such applications benefit from edge and fog computing to improve responsiveness and minimize the overhead of data exchange with cloud platforms \cite{ietf_edge}.
For example, image processing through artificial intelligence on IoT devices presents several benefits including faster decision-making and lower reaction time to environmental changes, less wireless interference with nearby devices, and improved security due to eliminating the need to upload raw images to a cloud service \cite{Griffiths_2018}.
According to ABI Research \cite{kelly2016internet}, 90\% of the data generated by edge devices is being processed locally.
In summary, local storage and processing can help satisfy the stringent latency requirements of mission-critical applications \cite{dezfouli2017rewimo}, reduces network utilization, reduces the processing overhead of resource-constraint devices, enhances security, and enables the system to continue its operation even in the presence of intermittent network connectivity \cite{chiang2016fog}.

Among the most-popular OSs for IoT edge devices is Linux.
According to the Eclipse IoT survey report \cite{iot_survey}, 71\% of IoT developers rely on this OS.
Existing low-cost devices such as the Raspberry Pi 3 (RPi3), Raspberry Pi Zero Wireless (RPiZW), Arduino Yun, and Beaglebone Black support this OS.
Despite low-level software and hardware improvements targeted at lowering their energy consumption, widespread use of these devices in IoT contexts requires employing user-level energy conservation techniques. 
Unfortunately, although there are valuable studies on the power measurement and modeling of this device \cite{kaup2014powerpi,morabito2017virtualization,vujovic2014raspberry,fisher2015open}, less attention has been paid to improving energy efficiency.

One of the most effective approaches towards improving energy efficiency is \textit{duty cycling}.
This approach has been widely employed by the wireless sensor network community to achieve a long node lifetime, in some cases up to a few years \cite{dezfouli2017rewimo,dezfouli2015modeling,dezfouli2015dicsa}.
In a duty-cycled application, a device powers on, performs its intended task, and then powers off for a specified interval.
Without relying on duty cycling, the lifetime of a system using an RPi3 and a 2400mAh battery is around 6 hours, assuming 400mA and 5V for current and voltage consumption.

\begin{figure}[!t]
\centering
    \includegraphics[width=1\linewidth]{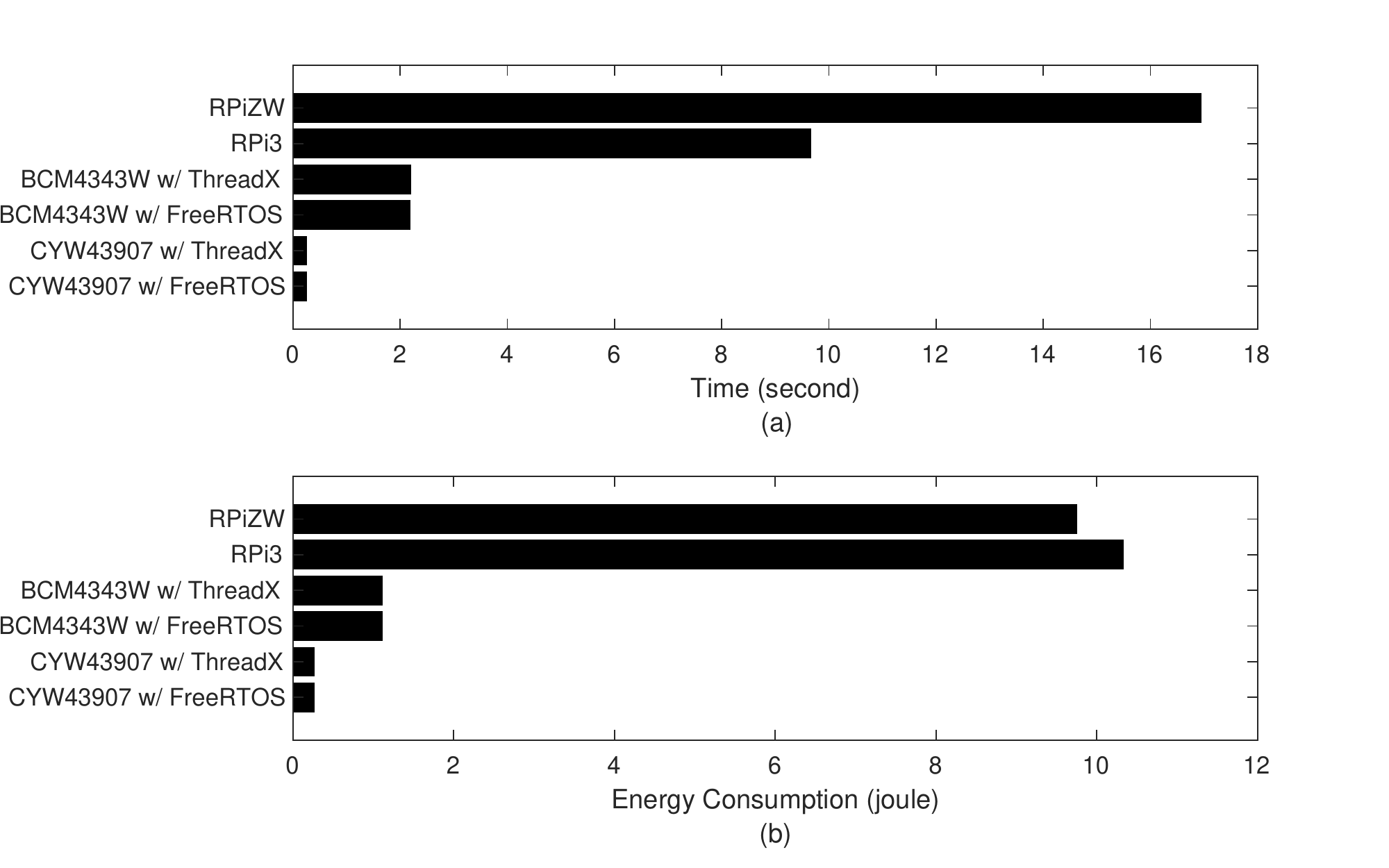}
    \caption{The (a) boot up duration and (b) energy consumption of the popular IoT operating systems on different hardware platforms.}
    \label{fig:boot_time_compare}
\end{figure}

The major burden of a duty-cycled Linux-based device is its boot up time.
Specifically, the system software and hardware must be loaded and initialized before running user applications.
Compared to real-time OSs such as FreeRTOS \cite{FreeRTOS} and ThreadX \cite{ThreadX}, the boot up time of Linux is several orders of magnitude longer.
To show this, we have measured the boot up time of various hardware platforms when using Linux, FreeRTOS and ThreadX.
The results are presented in Figure \ref{fig:boot_time_compare}.
These results indicate that for Linux-based systems a significant amount of energy is consumed during the boot up process at the beginning of each duty cycle.

There are two types of energy optimization techniques commonly applied to reduce the energy consumption of Linux:  
\textit{general optimization} \cite{chung2007study,googleDebian,kaminaga2006improving,Villegas2006ImproveTD, Bird2010MethodsTI} and \textit{application-specific optimization} \cite{singh2011optimizing, jo2009optimizing}.
General optimization refers to the improvement of OS code to run faster and consume less energy, agnostic to the application type.
Sample techniques belonging to this category include improving the filesystem and compression methods of unpacking the kernel during the bootloader phase \cite{chung2007study}, running boot up scripts in parallel and disabling kernel print statements \cite{googleDebian}, and saving a copy of the boot image to a file for reuse on subsequent boot ups \cite{kaminaga2006improving}.
Applying general optimization techniques, however, requires a deep understanding of the boot up process and OS.
In addition, hardware initialization may pose some challenges. 
For example, assume that a system image is taken after the hardware devices have been initialized.
When this image is reloaded during the next duty cycle, some hardware devices might not be initialized, and therefore the user application cannot function properly.
Addressing this limitation requires deep system knowledge and thorough testing to ensure system reliability.
Additionally, applying security updates requires image regeneration.
Even if the image can be updated, reliability would be a major concern. 
If images are generated on-the-fly after updates, the system might fail to boot up if the image is corrupted. 
Enabling fallback images requires an overhead for image verification and uses at least twice the storage space.

Application-specific optimization, on the other hand, refers to either OS improvement or the removal of unnecessary components, depending on application requirements.
For example, by customizing the bootloader and thinning the kernel of unnecessary modules, the authors of \cite{singh2011optimizing} decreased the boot up time on an embedded Android device by 65\%.
In \cite{jo2009optimizing}, the authors optimized a Linux-based smart television and reduced its boot up time by five seconds.
However, since they defined boot up time as the interval between power on and the time instance at which the user can interact with the device, they delayed the initialization of certain components until after the home screen of the television was loaded.
Unfortunately, the existing application-specific optimization techniques do not provide simple and universally applicable guidelines for IoT scenarios.
When Linux-based devices are used in IoT applications, it is desirable to tailor the system based on application requirements.
Even the Linux distributions released for IoT boards are preloaded with unnecessary services and initialize hardware devices not required by specific IoT application scenarios.
For example, a device running an image classification algorithm may not require sound utilities, remote login, service discovery daemons, time synchronization, or all the wireless technologies offered by the device.

In this paper we focus on the userspace level and application-specific optimization to improve the performance of duty-cycled Linux systems.
In other words, the goal of this paper is to profile and improve the energy consumed by the boot up and shutdown phases of the RPi3 (Raspberry Pi 3, based on the quad-core BCM2837 SoC) and the RPiZW (Raspberry Pi Zero w/ Wireless, based on a single-core BCM2835 SoC) in such a way that minimal work is required to tailor a standardized installation image to specific IoT applications.
In particular, the contributions of this paper are as follows:
First, we overview the boot up process and present the implementation of a testbed to measure the duration and energy consumption of this process.
In order to reveal the effect of loading services on system performance, we profile the start and end of loading units that require more than 10ms to initialize.
By categorizing system units into multiple classes, we show that customizing the set of active units, which we refer to as the \textit{unit configuration,} is a very effective approach towards improving duty-cycling performance.
For example, unit configuration reduces the energy consumption of the boot up process using a RPi3 by 43.62\% for an application that only requires communication with the camera interface. 
In addition, through profiling the resource utilization of the RPi3 and RPiZW in terms of processing, memory and I/O, we further analyze the overhead of initializing units and show the possibility of running user application processes while userspace initialization is still in progress.
Second, in the context of flash memory, even when up to 95\% of its capacity is utilized, our results show no effect on the boot up duration or energy consumption.  
However, using faster flash memory always results in a slightly lower boot up duration (around 1.5\%) and energy consumption (around 2.5\%).
Third, we investigate two shutdown approaches, \textit{graceful} and \textit{forced}, and evaluate the impact of unit configuration on their performance.
Our results confirm that using unit configuration reduces the energy consumption of the graceful and forced shutdown by up to 43.87\% and 57.42\%, respectively.
Additionally, the benefits and risks of both categories are highlighted when used for duty-cycled systems.
Fourth, we propose \textit{Pallex}, a parallel execution framework to execute a user application while userspace initialization is still in progress.
Using Pallex, a user application is split into several stages that execute at different points of the userspace initialization phase.
Our evaluations considering different user application scenarios show that in terms of lifetime, Pallex improves the duty-cycling performance of the RPi3 and RPiZW by 30.2\% and 9\%, respectively.
Although the power consumption of the RPi3 (quad-core) is higher than that of the RPiZW (single-core), the RPi3 achieves a longer lifetime due to its significantly shorter processing duration.

We have chosen to conduct our research using Raspbian Stretch Lite (RSL) on both the RPi3 and RPiZW because around 43\% of Linux-based IoT systems rely on Raspbian \cite{iot_survey}.
RSL is a popular release packaged without a desktop environment, is advertised as a minimalist distribution, has a long development cycle, and is officially supported by the Raspberry Pi Foundation, thereby making it an ideal candidate for deploying energy-efficient IoT applications.
However, it is worth noting that the results of this paper translate over to other Linux distributions supporting \texttt{systemd} such as Debian, Arch Linux, and Kali.

The rest of this paper is organized as follows:  
In Section \ref{section::bootProcess}, we provide an in-depth description of the Linux boot up process.
We present the components of our testbed and experimentation methodology in Section \ref{section:method}.
In Section \ref{section::performanceEnhancement}, we profile the Linux boot up process and measure the effect of unit configuration and flash memory on the duration and energy consumption of boot up.
The Linux shutdown phase is studied in Section \ref{section:sys_shutdown}.
The operation and performance evaluation of Pallex is presented Section \ref{section:PEF}.
Section \ref{section::related} overviews the related work.
We conclude our findings in Section \ref{section::conclusion}.

\begin{table}[!t]
\scriptsize
\centering
\caption{Key abbreviations and notations}
\label{Table_Abbreviations}
\begin{tabular}{p{0.13\linewidth}@{}| p{0.7\linewidth}@{}}
\hline
$E_{btl}$ & Energy of the bootloader phase \\
$E_{knl}$ & Energy of the kernel phase \\
$E_{usi}$ & Energy of the userspace initialization phase \\
$E_{sdn}$ & Energy of the shutdown phase \\
EU & Essential Units\\
GPIO & General Purpose Input/Output\\
NRS & Networking Related Services \\
$P_{btl}$ & Bootloader phase \\
$P_{knl}$ & Kernel phase \\
$P_{usi}$ & Userspace initialization phase \\
RPi & Raspberry Pi 3 or  Raspberry Pi Zero w/ Wireless\\
RPi3 & Raspberry Pi 3 \\
RSL & Raspbian Stretch Lite \\
RPiZW & Raspberry Pi Zero w/ Wireless \\
SDC & Secure Digital Card \\
SoC & System on a Chip \\
$T_{btl}$ & Duration of the bootloader phase \\
$T_{knl}$ & Duration of the kernel phase \\
$T_{usi}$ & Duration of the userspace initialization phase \\
$T_{sdn}$ & Duration of the shutdown phase \\
UART & Universal Asynchronous Receiver Transmitter
\\
\hline
\end{tabular}
\end{table}


\section{Boot Up Process}
\label{section::bootProcess}

\begin{figure}[!t]
\centering
    \includegraphics[width=0.99\linewidth]{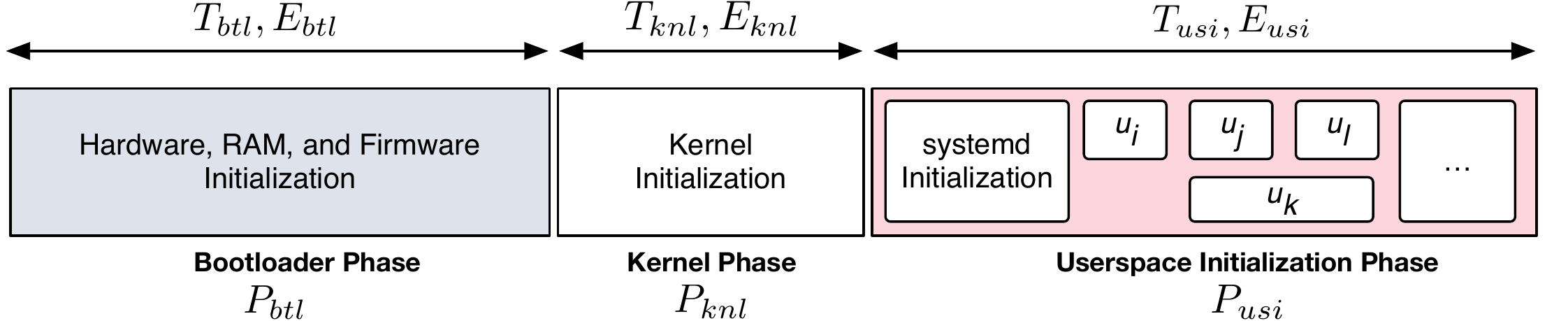}
    \caption{The sequence of operations during the boot up process. $u_{i}$, $u_{j}$, $u_{k}$ and $u_{l}$ refer to userspace units.}
    \label{fig:boot_steps}
\end{figure}

As demonstrated in Figure \ref{fig:boot_steps}, the Linux boot up process consists of three main phases: (i) the bootloader phase ($P_{btl}$), (ii) the kernel phase ($P_{knl}$), and (iii) the userspace initialization phase ($P_{usi}$).  
Each of these phases present unique opportunities to optimize boot time.  
In this paper we focus on $P_{usi}$ because it enables the user to simply and effectively implement duty cycling to conserve energy.

\subsection{Bootloader Phase}
In the Pi's boot up process, a two-stage bootloader\footnote{Until October 2012, 
the RPi platform used a three stage bootloader, with an additional file, \texttt{loader.bin}, executed by the GPU between the \texttt{bootcode.bin} and \texttt{start.elf} stages.} prepares the hardware to load the kernel. 
First, while the ARM processor is off, the GPU is powered up and initializes itself by executing a first stage bootloader that is burned into the SoC's ROM \cite{upton2016learning}. 
This stage instructs the GPU to power on the Secure Digital Card (SDC) and read a file called \texttt{bootcode.bin} from the first partition of the SDC. 
The execution of \texttt{bootcode.bin} enables the on-board SDRAM and loads \texttt{start.elf}, which contains firmware for the GPU. 
After reading in system configuration parameters from \texttt{config.txt}, the GPU loads the kernel image (\texttt{kernel.img}) along with kernel parameters (\texttt{cmdline.txt}) into the shared RAM allocated to the ARM processor. 
Lastly, the second stage powers on the ARM processor by triggering the reset signal \cite{BroadcomSpecSheet}.
Now the system is running the Linux kernel. 
In this paper we refer to the duration and energy consumption of this phase as $T_{btl}$ and $E_{btl}$, respectively.
Also, we refer to the ARM processor as "processor".

\subsection{Kernel Phase}
The Linux kernel handles all OS-related processes such as memory management, process scheduling, driver initialization, and overall system control. 
The kernel is initialized in two steps: 
The first step occurs when the kernel is loaded into memory and decompressed.
Basic memory management is enabled during this stage as well.
The next major step the kernel takes is launching an \texttt{init} process to run and transition to $P_{usi}$.
In this paper we refer to the duration and energy consumption of the kernel phase as $T_{knl}$ and $E_{knl}$, respectively.

\subsection{Userspace Initialization Phase}
The final phase of the boot up process is userspace initialization.
In this paper we refer to the duration and energy consumption of this phase as $T_{usi}$ and $E_{usi}$, respectively.
During this phase, the units that are activated (a.k.a., initialized) in the userspace as well as the daemons that run in the background during active mode are activated by an \texttt{init} process.
Most modern Linux distributions (including RSL) use \texttt{systemd} \cite{systemdInit,Gorauskas_2015,systemd_unit} as their initialization system.
The predecessor to \texttt{systemd} was \texttt{System V init} (\texttt{sysvinit}), which traces its origins back to the original commercial Unix system.
Compared to \texttt{sysvinit}, \texttt{systemd} offers advantages such as calendar-based job timers, a more unified API, and backward compatibility with \texttt{sysvinit}. 

\texttt{systemd} is the first daemon to start during boot up and the last to exit during shutdown. 
In addition to operating processes and services, \texttt{systemd} is capable of triggering filesystem mounts, monitoring network sockets and running timers.
Each of these capabilities is described by a set of configurations files, termed  \texttt{unit} files.
Unit types include service units, which manage background services; mount units, to mount filesystems; and target units: to group and control other units.
There are other unit types as well, but the details of their implementation reside outside the scope of this paper.
In order to manage the dependencies and ordering of  unit activation, unit files employ the syntax provided by \texttt{systemd}.
When \texttt{systemd} is initialized, it first loads the unit configurations and determines the boot up goal.
Based on the specified hierarchy of dependencies, \texttt{systemd} activates the units in order to reach the target goal.
On RSL, the \texttt{rc.local} file is loaded and executed by \texttt{systemd} after all the services have been initialized.
Therefore, the Raspberry Pi Foundation recommends general consumers launch user processes through this file to ensure all necessary hardware and software components are initialized.
However, performance improvements are easily achieved through manually resolving dependencies and writing custom unit files, as we will demonstrate in Section \ref{section::performanceEnhancement}.

One important feature \texttt{systemd} provides is the ability to activate units in parallel. 
This feature can save time compared to initializing units sequentially, even on a single-core or single-threaded board, as some units require time for hardware initialization \cite{LinuxKernelDev}.
Finally, \texttt{systemd} enables developers to customize rules for automatically starting, reloading, and killing services.




\section{Testbed Overview}
\label{section:method}

\begin{figure}[!t]
\centering
    \includegraphics[width=1\linewidth]{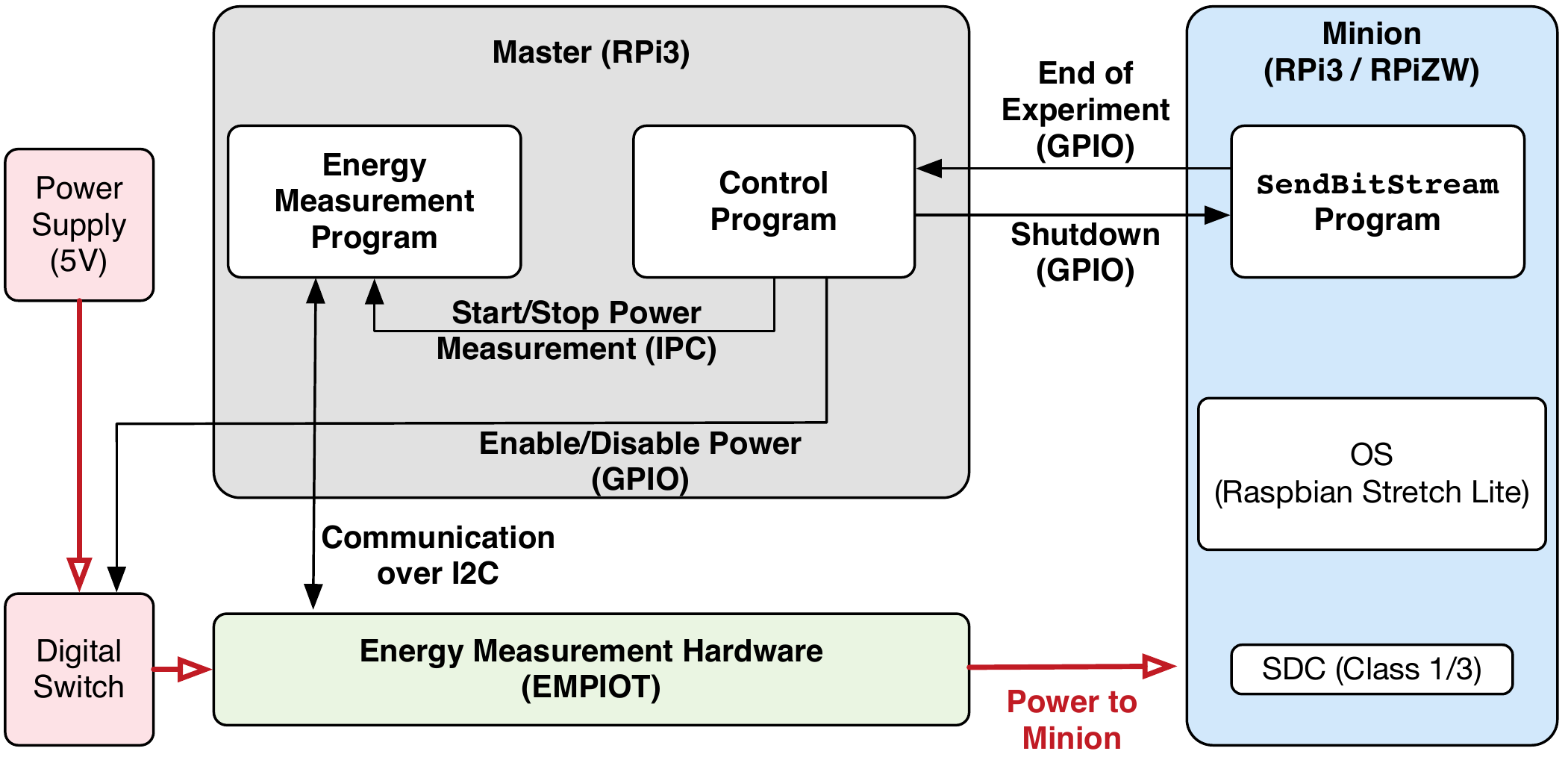}
    \caption{The schematic of the testbed used for measuring the duration and energy consumption of the boot up phases.}
    \label{testbed_fig}
\end{figure}

Figure \ref{testbed_fig} shows the architecture of the testbed.
It consists primarily of two hardware components: a \textit{master} and a \textit{minion}.
The master is composed of a RPi3 and uses a shield board \cite{dezfouli2018empiot} for energy measurement.
The minion refers to the device under test.
We used two different minion boards in this paper, a RPi3 and a RPiZW, where both run the March 2018 release of RSL.

\begin{figure}[!t]
\centering
    \includegraphics[width=0.7\linewidth]{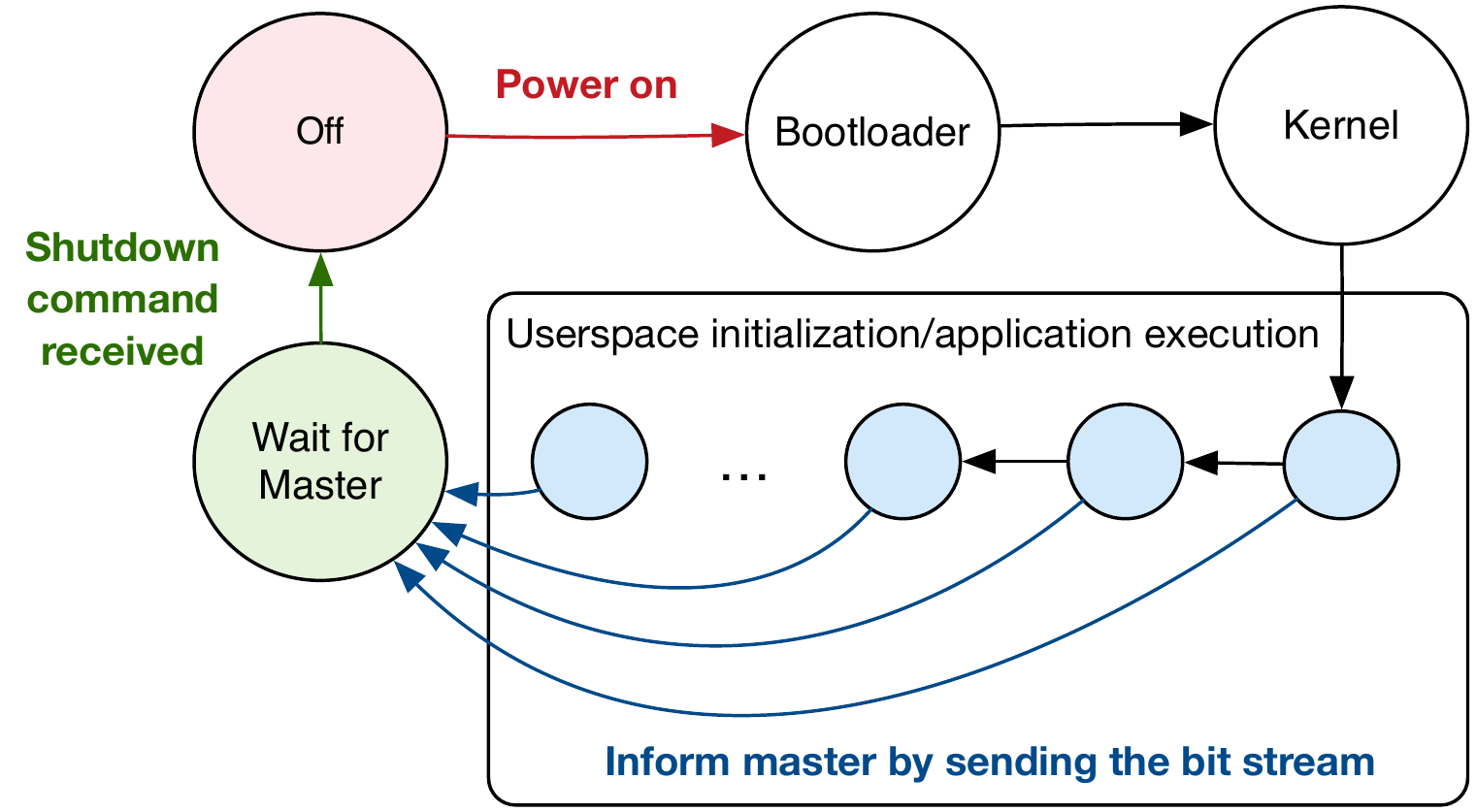}
    \caption{The state machine of the minion device during an experiment. The minion executes the \texttt{SendBitStream} program to send a message to the master after reaching the desired level of completion.}
    \label{fig:minion_steps}
\end{figure}

The master runs two programs: 
(i) a \textit{control program} that is responsible for enabling and disabling the input power to the minion, and 
(ii) an \textit{energy measurement program} that controls the EMPIOT shield to measure the energy consumption of the minion.
Algorithm \ref{alg:masterPseudocode} shows the pseudo-code of the control program running on the master device.
This program is used to control power supply to the minion and measure the duration and energy consumption of the boot up process by sending commands to the energy measurement program.
First, the BCM \cite{bcmdriver} and WiringPi \cite{wiringpi} libraries are initialized.
Then, the energy measurement program is initialized to listen on a socket and receive commands from the control program using inter-process communication (IPC).
At the beginning of each experiment, the control program communicates with the digital switch through GPIO pins controlled by the BCM library to turn on the minion. 
At the same time, the control program communicates with the energy measurement program through the socket to start power measurement.
After the start of the boot up process, the control program uses the WiringPi library to detect and decode a message sent by the minion when the experiment reaches the desired completion point.
The end of the operation depends on the experimentation scenario and refers to the cases such as the end of $P_{usi}$ or the completion of a user application.
The control program logs the duration and energy consumption of this operation for each experiment.
Figure \ref{fig:minion_steps} shows the states of the minion during one single experiment from the instance the minion is powered on until the instance the minion sends the signal to the master after it has completed the desired set of operations.

It should be noted that the minion cannot use Ethernet, WiFi, or Bluetooth to inform the master about the end of its operation because their relevant units might not be activated in some scenarios, as we will see later in this paper.
Furthermore, it is not possible to use any of these mechanisms when the performance of $P_{btl}$ and $P_{knl}$ is being measured.
To address this challenge, we send a message over a GPIO pin from the minion to the master.
When the minion completes its operation, it runs a program that generates a simple bit pattern to notify the master.
The minion generates a bit pattern, instead of a simple rising or falling edge signal, because of the GPIO voltage variations during the boot up process.
Therefore, the generated bit pattern avoids the master from reporting false positives.
This approach enables us to measure the performance of $P_{btl}$ and $P_{knl}$ because \texttt{systemd} is initialized immediately after the kernel phase, and GPIOs can be used as soon as \texttt{systemd} is initialized.
The pattern generation program, named \texttt{SendBitStream}, is a unit that is activated by \texttt{systemd} once the targeted state (based on the experiment type) has been reached.
Specifically, to measure $T_{btl} + T_{knl}$ (and $E_{btl} + E_{knl}$), this unit is called immediately after \texttt{systemd} initialization.
This is achieved by creating a new unit without enforcing any dependencies.
Similarly, to measure $T_{btl} + T_{knl} + T_{usi}$, this unit is called when all the required units have been activated.
In order to measure boot up duration and energy until a particular unit has been activated, this unit is activated when its dependencies have been resolved.
We used \texttt{systemd-analyze} to extract $T_{knl}$.
By using this value, we can compute $T_{btl}$ as well.
A similar approach is used to measure the energy consumption of these phases.


The existing COTS energy measurement tools are either costly or do not offer the necessary features to build a fully controllable energy measurement testbed. 
For example, Keithley 7510 \cite{kth_dmm_7510}, includes only 2MB of storage and costs more than \$3500.
The Monsoon tool, which has been widely used by the academia, interfaces with the board under test using a USB connection. 
Therefore, accurately controlling the start and stop of the measurement interval is not possible.
On the other hand, the tools proposed by the research community reveal the following shortcomings: (i) complexity and the need to modify the board under test (e.g., \cite{Jiang2007,Stathopoulos2008,Duttac2008,Andersen2009,Zhou2013a,Naderiparizi2016}), (ii) offering a limited measurement range (e.g., \cite{Haratcherev2008,Trathnigg2008,Zhu2013b,Hartung2016,Jiang2007,Potsch2017}), and (iii) low accuracy (e.g., \cite{Gomez2012a,Keranidis2014b}).
Due to these concerns, to build the testbed required for the experiments of this paper, we used the EMPIOT tool proposed in \cite{dezfouli2018empiot}.

EMPIOT is a shield board that is installed on top of an RPi3 and communicates with the energy measurement program using I2C, as Figure \ref{testbed_fig} shows.
For the experiments of this paper, we have configured the board to measure current and voltage values up to 1A and 5V, respectively.
For current measurement, the board utilizes the INA219 chip to measure the voltage across a 0.1$\Omega$ shunt resistor.
This voltage is then converted to current using Ohm's law.
The voltage is directly measured using the internal ADC of INA219.
The EMPIOT tool is capable of supersampling approximately 500,000 samples per second (sps), which are then averaged and streamed at 1Ksps.
The current and voltage resolution of this platform are 100$\mu$A and 4mV, respectively, when the 12-bit resolution mode is configured.
Once a new sample is ready, the shield sets the conversion ready bit, which is read by the energy measurement program running on the master.
The sample is then read off the shield board using I2C communication.
The energy measurement program uses two buffers.
Once a buffer it filled, an execution thread is activated to write the collected samples to a file, while the second buffer is being filled up.
This program also converts the raw power samples collected during the measurement interval to energy using Riemann integration.

\begin{algorithm}[t]
    	\footnotesize
	\SetInd{0.9em}{0.7em}
    
    \SetKwFunction{FMain}{startExperiments}
    \SetKwProg{Fn}{function}{}{}
    \Fn{\FMain{}}
    {
        \texttt{\textcolor{algCommColor} {/*\textit{to control the power switch through a GPIO pin and receive the bit pattern from the minion on another GPIO pin} */}}\\
        setup the BCM and WiringPi GPIO libraries\;
        \BlankLine 
        \texttt{\textcolor{algCommColor} {/*\textit{setup the energy measurement program and make it ready to measure power} */}}\\
        initialize the energy measurement energy\;
        \BlankLine 
        \For{i = 0; i < 50; i++} 
        {
            cut power to the minion\;
            apply power to the minion and start time and energy measurement\;
            wait for minion to transmit bit pattern\;
            record duration and energy consumption\;
        }
    } 
    
    \caption{Master's control program}
    
    \label{alg:masterPseudocode}

\end{algorithm}

\begin{figure}[!t]
\centering
    \includegraphics[width=0.6\linewidth]{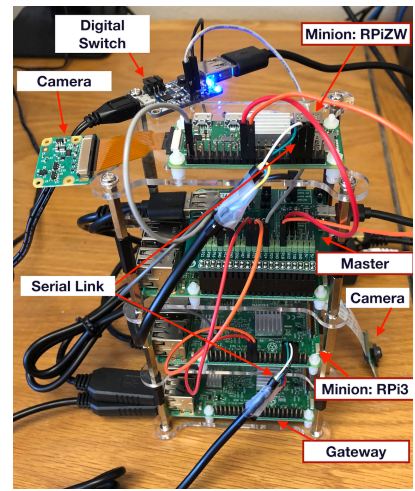}
    \caption{The hardware components of the testbed.}
    \label{testbed_actual}
\end{figure}

Figure \ref{testbed_actual} shows the actual testbed used.
In addition to the master node and the two minion boards, this testbed includes a gateway, which is a RPi3 board.
Since for most of the scenarios, both the wired and wireless communication interfaces are disabled, we need to use the serial port to communicate with the minions and configure the tests.
To this end, the gateway, which is always connected to our wired network, enables us to use UART and communicate with the minions.
For the rest of this paper we refer to a minion board simply as "board", which is the device under test.

Unless otherwise mentioned, results are averaged over 50 experiments, with the error bars representing the 95\% confidence interval of the mean.

\section{Profiling and Enhancement of the Linux Boot Up Process}
\label{section::performanceEnhancement}

In this section, we first study the activation time of units during $P_{usi}$.
We then profile system resource utilization in terms of processing, memory and I/O.
In addition, we evaluate the effect of SDC speed and capacity usage on the duration and energy consumption of the three phases of the boot up process.
Finally, we show how customizing userspace units can be employed to improve boot up duration and energy consumption.


\subsection{A Deeper Look into the Userspace Initialization Phase}

The userspace initialization phase activates a variety of units supporting different functionalities of the system. 
A summary of these units is available in Appendix \ref{appendix:services}.
In the beginning of this phase, \texttt{systemd} is initialized and it loads unit configuration files to determine which units must be activated.
It then creates a dependency tree to determine the ordering of unit dependency resolution.
In order to improve efficiency, units can be initialized in parallel with respect to their dependency relationships.

\begin{figure*}[t]
\centering
    \includegraphics[width=1\linewidth]{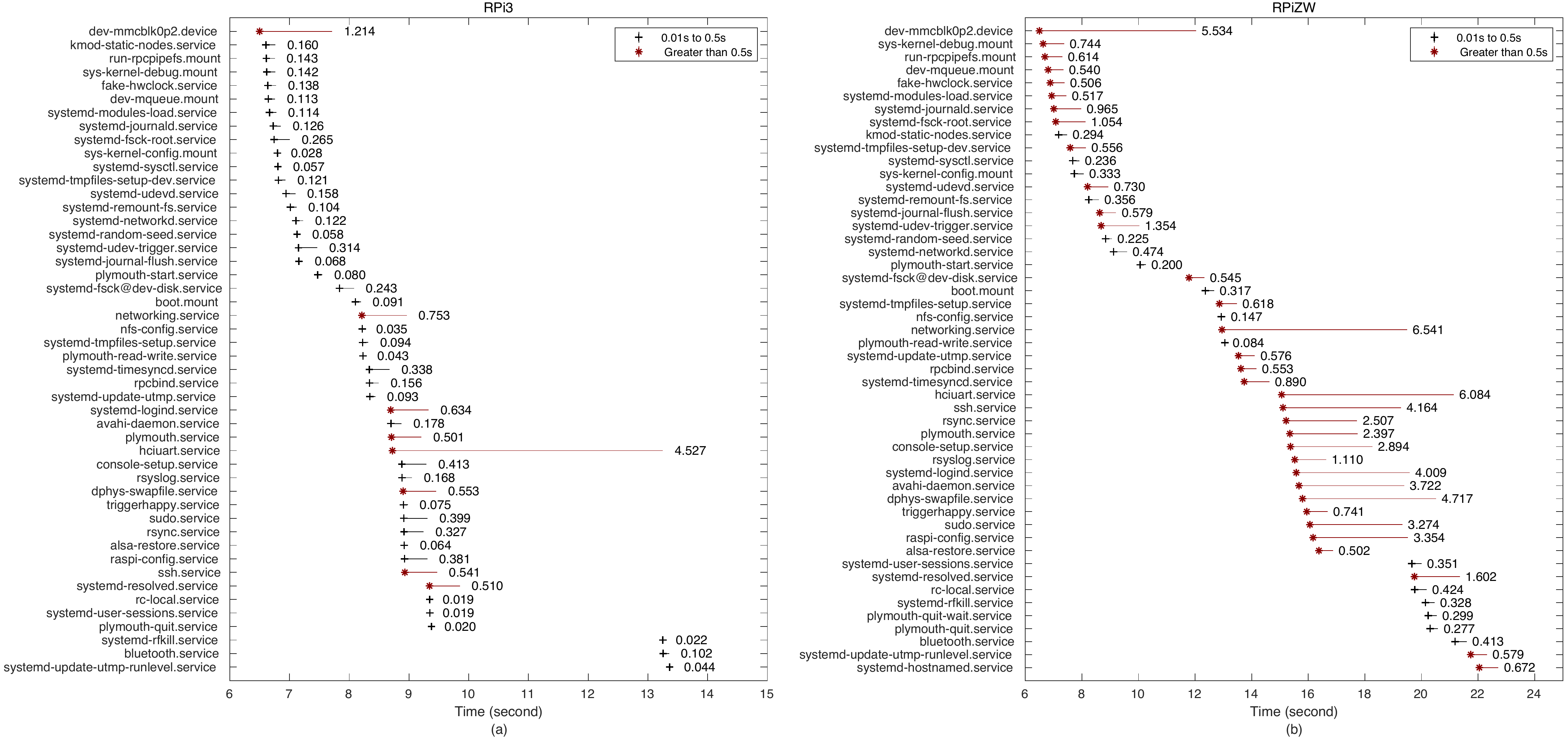}
    \caption{The starting time and initialization duration of units during $P_{usi}$ for (a) RPi3 and (b) RPiZW.  
     Only the units with activation duration longer than 10ms are included in this figure. Results are averaged over 50 experiments.}
    \label{fig:gannt}
\end{figure*}

Figures \ref{fig:gannt}(a) and (b) show the start and activation duration of units during $P_{usi}$ for the RPi3 and RPiZW, respectively.
We have used the \texttt{systemd-analyze blame} utility to extract these data.
Although all of the units are enabled for these experiments, we disabled WiFi and Ethernet connectivity to extract the activation duration of units without them being affected by external factors such as communication with a WiFi access point.
Also, to focus on the units that significantly contribute to $T_{usi}$, these figures do not display units that require less than 10ms to be activated.
Please note that the x-axis of both Figures \ref{fig:gannt}(a) and (b) start at $t = 6s$ because $T_{btl} = 3.65s$ and $T_{knl} =  2.85s$ for both boards.
It must also be noted that, since $energy = power \times duration$, the execution duration of each unit does not necessarily reflect its energy consumption. 
We will study the energy consumption of unit activation in the subsequent sections.
Additionally, we will show that a long unit activation duration does not prevent us from running a user application concurrently with the unit activation.

%
%
%
%
%
%


For the RPi3, our results show that the system units imposing the highest overhead are \texttt{dev-mmcblk0p2.device}, \texttt{networking.service}, \texttt{hciuart.service}, and \texttt{systemd-resolved.service}.
Except the first unit, which is responsible for bringing the root partition into the scope of \texttt{systemd}, the rest are networking-related services.
Networking services generally require a longer activation duration compared to other units because they require a combination of initializing hardware and networking utilities. 
The \texttt{hciuart.service} is responsible for the initialization of Host Controller Interface (HCI) to provide a uniform interface for accessing Bluetooth hardware capabilities.
All USB-Bluetooth adapters operate with a HCI interface over the USB link.
During its initialization, HCI creates read and write communication threads, establishes a connection to the Bluetooth transceiver, and reads device buffer sizes.
Enabling real-time Bluetooth communication requires the processor to frequently context-switch to monitor UART communications. 
Because \texttt{bluetooth.service} depends on \texttt{hciuart.service}, it finishes initialization after \texttt{hciuart.service}.
An interesting behavior of these services is that they both complete their activation after \texttt{rc.local}.
This means that user applications that rely on Bluetooth cannot be started using \texttt{rc.local}.
Instead, these applications require a \texttt{systemd} unit that is scheduled to be activated after the completion of \texttt{bluetooth.service}.


On the RPiZW, we observe that $T_{usi}$ is about 17s, which is 10s longer than that of the RPi3.
The same units mentioned for the RPi3 consume most of the userspace initialization time on the RPiZW as well.
Since the RPiZW's processor only has one core, compared to the RPi3's quad-core processor, the difference in duration is expected.
A multi-core processor can parallelize unit activations across multiple cores, while a single-core processor must context-switch more frequently between tasks.
This also explains why different sets of units are presented in Figure \ref{fig:gannt}(a) and (b).
Furthermore, other units including \texttt{systemd-login.service}, \texttt{console-setup.service}, \texttt{alsa-restore.service}, and \texttt{systemd-user-sessions.service} reportedly require a longer activation duration (at least 1s) on the RPiZW.

\textbf{Threats to Validity.}
Our analysis of unit activation duration revealed one significant shortcoming about \texttt{systemd-analyze blame}. 
The initialization duration calculated by this utility is not completely accurate if the dependency tree is not precisely configured.
For example, if \texttt{systemd} attempts to activate a fast unit that depends on a longer unit which has not been activated yet, the faster unit cannot complete its activation until the dependency has been resolved.
More specifically, if the activation duration of longer and shorter units are 800ms and 10ms, respectively, then \texttt{systemd-analyze blame} reports 810ms as the initialization duration of the faster unit.
For example, for \texttt{sudo.service}, \texttt{systemd-analyze blame} falsely reports a long initialization duration because \texttt{systemd} attempts to activate it before the filesystem the service depends on is mounted.
Therefore, the values generated by this utility may be longer than the actual activation duration of each unit, but these values are not shorter.
To cover all the units that contribute significantly to the boot up process, we have included only those units that require more than 10ms according to \texttt{systemd-analyze blame}.
In Figure \ref{fig:gannt}, we include Type I errors to avoid excluding any units that require more than 10ms. 
Later, to eliminate Type I errors, we identify and study the impact of units that significantly affect the boot up process.

\subsection{Customizing Userspace Initialization}
\label{customize_units}

In this section, we analyze the effect of disabling optional units on the performance of the userspace initialization phase.
This is referred to as \textit{unit configuration} in this paper.
The \texttt{systemctl} \cite{systemctlService} utility is used to implement unit configuration.

Among the units activated during $P_{usi}$, some of them are essential to maintain stable system operation.
For example, units that are responsible for mounting the file systems and loading kernel modules cannot be safely disabled.
Another example, nearly all user applications will work even if \texttt{systemd-random-seed.service} is disabled.
However, disabling this service is a security risk, because it is critical for maintaining higher entropy for the secure generation of random numbers used in encryption algorithms.
A complete list of these units, which are referred to as \textit{Essential Units} (EU) in this paper, can be found in Appendix \ref{app_es}. 
The next category includes a significant number of services and is referred to as \textit{Networking-Related Services} (NRS) in this paper.
Appendix \ref{appendix_nrs} overviews these services.
These services are by far the most variable in terms of activation duration because they often rely on external dependencies (e.g., association with an access point, communicating with a DHCP server, etc.) and/or initializing physical hardware such as the WiFi and Bluetooth transceivers.
Due to the significant effect of NRS on boot up performance, we study the effect of following unit configurations on $P_{usi}$:

\begin{itemize}
    \item \textit{EU}. Refers to the case where only the essential units (cf. Appendix \ref{app_es}) are enabled.
    For EU configuration, we detect the end of userspace initialization when \texttt{rc.local} runs because it is the last unit file that is executed.
    \item \textit{MMS}. Refers to \texttt{dphys-swapfile.service}.
    For configuration EU w/ MMS, we detect the end of userspace initialization when \texttt{rc.local} runs because it is the last unit file that is executed.
    \item \textit{NET1}. Refers to \texttt{networking.service}.
    For configuration EU w/ NET1, we detect the end of userspace initialization when \texttt{rc.local} runs because it is the last unit file that is executed.
    \item \textit{NET2}. Refers to \texttt{networking.service} and \texttt{sshd.service}.
    For configuration EU w/ NET2, we detect the end of userspace initialization when \texttt{rc.local} runs because it is the last unit file that is executed.
    \item \textit{NET3}. Refers to \texttt{bluetoothd.service} and \texttt{hciuart.service}.
    For configuration EU w/ NET3, since \texttt{bluetooth.service} is the last service that is executed, we detect the end of userspace initialization when this service completes its initialization.
    \item \textit{ALLU}. Refers to the case where all units are enabled.
    For this configuration, since \texttt{bluetooth.service} is the last service that completes, we detect the end of userspace initialization when this service is activated. 
    For configuration ALLU w/o NET3, we detect the end of userspace initialization when \texttt{rc.local} runs because it is the last unit file that is executed.
\end{itemize}

\begin{figure*}[t]
\centering
    \includegraphics[width=0.9\linewidth]{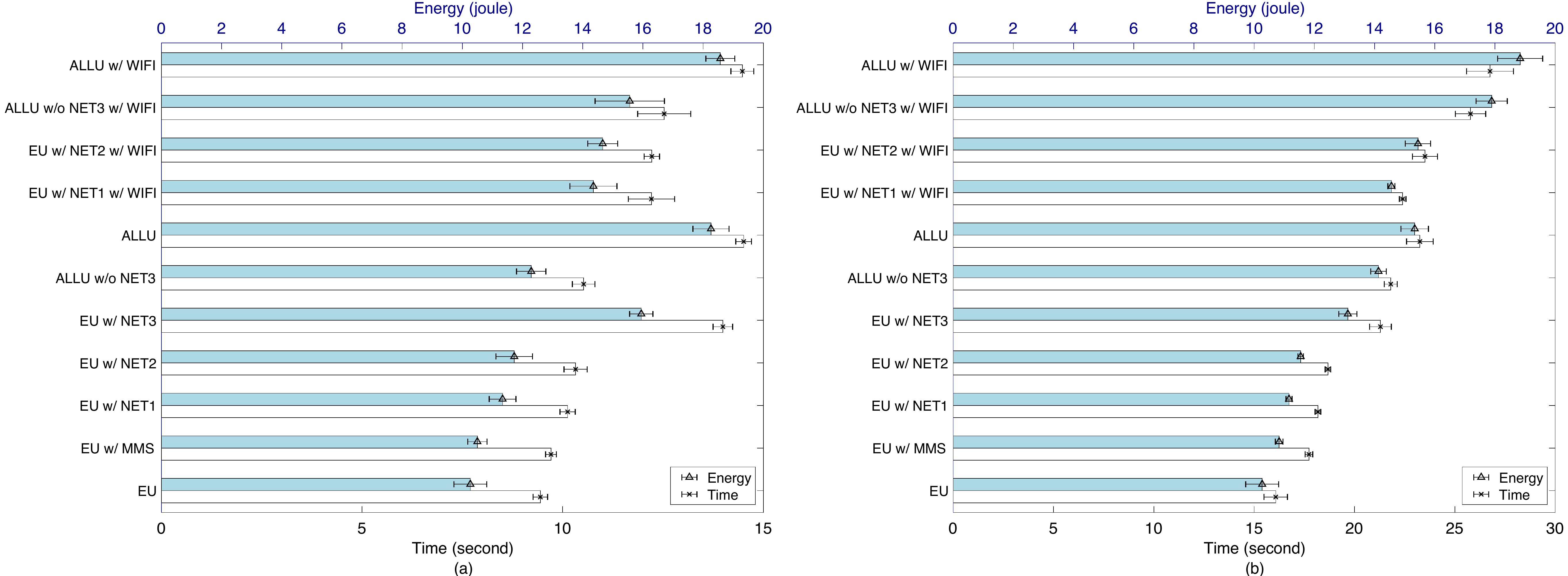}
    \caption{The duration and energy consumption of booting up the (a) RPi3 and (b) RPiZW for different unit configurations. The lower and upper axes show $T_{btl} + T_{knl} + T_{usi}$ and $E_{btl} + E_{knl} + E_{usi}$, respectively.}
    \label{fig:boot_time_energy_three_zero}
\end{figure*}

In addition to the classifications detailed above, since the actual association of a RPi with an access point introduces more variations due to the control messages exchanged between the two parties, we report the results separately for the cases where a WiFi connection is established.
Figures \ref{fig:boot_time_energy_three_zero}(a) and (b) show the impact of unit configuration on both the RPi3 and RPiZW, respectively, in terms of $T_{usi}$ and $E_{usi}$.

%


These figures clearly show the benefits of unit configuration to enhance the performance of boot up phase.
For example, applying unit configuration EU reduces the energy consumption by 43.62\% compared to ALLU, for the RPi3 board.
These results also reveal that WiFi significantly affects $T_{usi}$ and $E_{usi}$.
Specifically, for configuration EU w/NET1, enabling WiFi increases $T_{usi}$ by around 2s and 4s, for the RPi3 and RPiZW, respectively.
Besides, for this configuration, enabling WiFi increases $E_{usi}$ by 3 and 3.5 joules for the RPi3 and RPiZW, respectively.



These results also reveal the higher effect of \texttt{hciuart.service} and \texttt{bluetooth.service} on the RPi3.
Since activating these services do not significantly utilize the processing resources of the RPi3 (as we will show in Section \ref{rsc_util}), the waste of processing resources, and hence energy consumption, is higher than that of the RPiZW.
Therefore, EU w/ NET3 compared to EU results in a 54.8\% energy increase on RPi3, compared to the 26.92\% increase on RPiZW.
In addition, we can observe that on the RPi3, which can initialize the WiFi and NET3 services concurrently, the duration and energy consumption of the "ALLU" and "ALLU w/ WiFi" configurations are almost equal.
In contrast, for the RPiZW, the duration and energy consumption of the "ALLU w/ WiFi" configuration are higher than "ALLU" because the single-core processor needs to interleave the tasks of initializing WiFi and NET3 services.
We will further study this behaviour in the next section.

It must be noted that \textit{disabling} units does not necessarily prevent their activation by \texttt{systemd}.
More specifically, units might be activated when: (i) other units relying on them are activated, or (ii) in the case of external event hooks such as attaching a device.
For example, \texttt{alsa-restore.service} is automatically activated even if it has been disabled.
However, as its main function is to initialize the onboard sound card, it is not required by most IoT applications.
Therefore, it is worth \textit{masking} this service using \texttt{systemctl}.
This is performed by pointing the unit file to the special device \texttt{/dev/null} so that the dependency tree can mark it as resolved for dependants without actually running it.


The exact increase of duration and energy when using the networking services depends on factors such as interference, channel congestion, and the load of the access point during the association process.
For example, the duration of WPA authentication and IP allocation increases as the current load of the access point is intensified.
Since these external dependencies are outside the scope of the RPi performance, unnecessary services must be disabled or careful attention must be paid to link quality and access point load to achieve a desirable performance.

\subsection{Resource Utilization During Userspace Initialization}
\label{rsc_util}

\begin{figure}[t!]
\centering
    \includegraphics[width=1\linewidth]{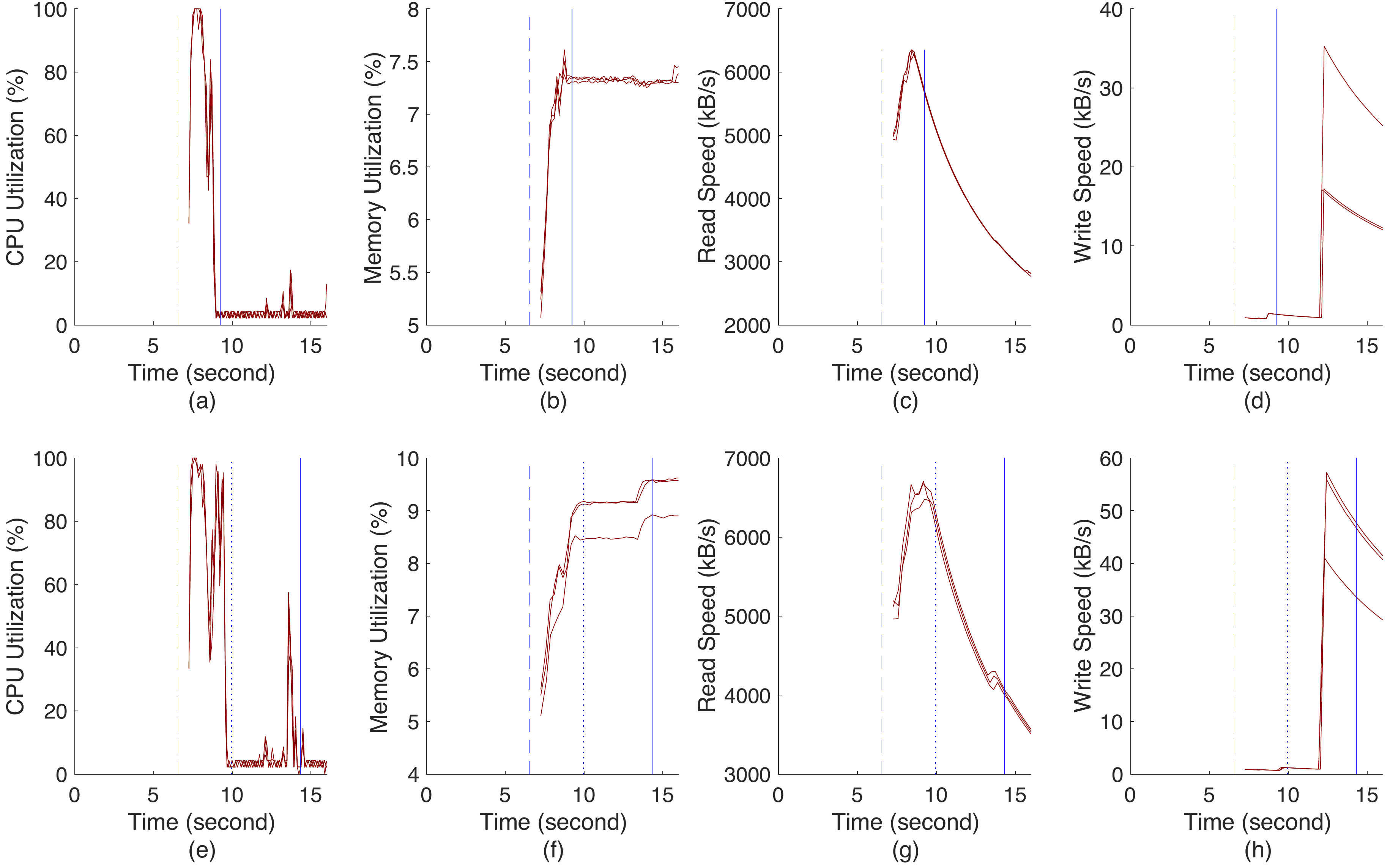}
    \caption{The resource utilization of the RPi3 during userspace initialization for three different experiments. 
    Sub-figures (a)-(d) represent unit configuration EU, and sub-figures (e)-(h) represent unit configuration ALLU.
    Dashed lines indicate the start of $P_{usi}$, dotted lines indicate the instance \texttt{rc.local} is activated, and solid lines indicate the end of $P_{usi}$.
    }
    \label{fig:cpu_three}
\end{figure}

\begin{figure}[!t]
\centering
    \includegraphics[width=1\linewidth]{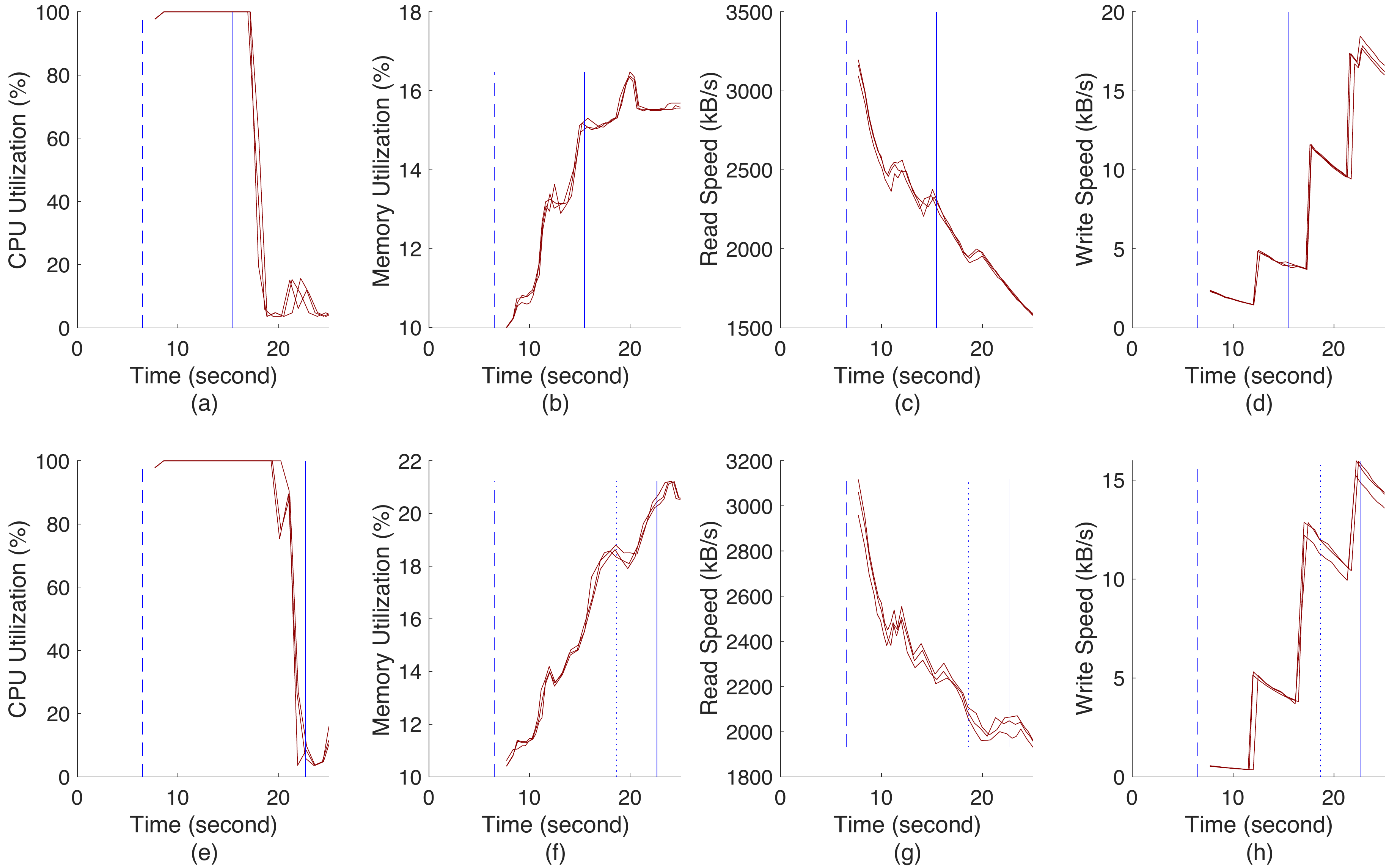}
    \caption{The resource utilization of the RPiZW during userspace initialization for 3 different experiments.    
    Sub-figures (a)-(d) represent unit configuration EU, and sub-figures (e)-(h) represent unit configuration ALLU.
    Dashed lines indicate the start of $P_{usi}$, dotted lines indicate the instance \texttt{rc.local} is activated, and solid lines indicate the end of $P_{usi}$.
    }
    \label{fig:cpu_zero}
\end{figure}

In this section, we study system resource utilization during $P_{usi}$.
To this end, we measured processor utilization, memory utilization, and SDC's I/O speed on the RPi3 and RPiZW.
Resource monitoring is performed by a shell program that starts as soon as \texttt{systemd} is initialized.
For this reason we noticed that resource monitoring is not available for $0.75$s and $1.2$s after the completion of $P_{knl}$ on the RPi3 and RPiZW, respectively.
To record processor utilization, we wrote a \texttt{gawk} \cite{gawk} script to read values from \texttt{/proc/stat} and calculate the current percentage of processor utilization across all cores with high granularity and low overhead.
We read directly from \texttt{/proc/meminfo} to determine memory utilization, and used the \texttt{iostat} \cite{iostat} utility for collecting SDC's I/O utilization.
Figures \ref{fig:cpu_three} and \ref{fig:cpu_zero} show the results for three trials.

%

%

Comparing the two figures indicates the significantly higher processor utilization of the RPiZW compared to the RPi3.
For the RPi3, we notice that processor utilization drops to less than 5\% as soon as \texttt{rc.local} is invoked, regardless of the unit configuration applied (compare Figure \ref{fig:cpu_three}(a) and (e)).
For the RPiZW, however, when the unit configuration ALLU is applied, processor utilization drops to around 5\% after the end of userspace initialization, which is the completion of NET3 services (compare Figures \ref{fig:cpu_zero}(a) and (e)).
We justify this behavior by referring back to Figure \ref{fig:gannt}.
While the RPi3 is almost finished with userspace initialization (except the NET3 services) at the time \texttt{rc.local} is activated, RPiZW needs to complete the activation of multiple units.
Specifically, as Figure \ref{fig:gannt} shows, a considerable number of units with loading time longer than 10ms are being activated around $t = 20$.

For RPi3, enabling all services increases memory utilization from around 7\% to 10\%.
For RPiZW, the increase is from around 16\% to 21\%.
These results indicate that even for the unit configuration ALLU, more than 900MB and 400MB of RAM is available on the RPi3 and RPiZW, respectively.
In terms of I/O, since the RPi3 initializes more units in parallel, its I/O speed is almost double than that of the RPiZW.
Although the processor utilization of the RPiZW is around 100\% throughout $P_{usi}$, we can still benefit from concurrent execution of user applications within this phase if they mostly rely on peripheral initialization and I/O operations.
A sample IoT application that satisfies this requirement is capturing a photo using a camera.
For example, the RPi camera module \cite{RPiCam} connects over a Camera Serial Interface (CSI) to the GPU on the RPi. 
While a picture is being captured and processed, the processor can switch to other tasks, as there are several steps performed by the camera that are independent of the RPi's processor.
In particular, the camera has a physical requirement of exposure time.
Next, the camera needs to process the image.
According to the manufacturer of the camera module's image sensor (IMX219PQ \cite{pi_camera_cmos}), the camera has Lens Shading Correction (LSC) functionality, which means the image undergoes some processing on the camera module before it is sent over CSI.
After this step, the image must be sent over the CSI interface as a series of Bayer frames to the RPi's GPU.
Next, the GPU's VideoCore firmware assembles the image.
Finally, the processor can receive the assembled image from the GPU.
In Section \ref{section:PEF} we will show the performance improvement of this application running during $P_{usi}$.


\subsection{Profiling the Time and Energy Consumption of Bootloader and Kernel Phases}
\label{profiling_energy_time_btl_knl}

In this section, we study the duration and energy consumption of bootloader, kernel, and userspace initialization phases versus the properties of the SDC used.
In order to measure the effect of SDC speed on performance, we used two 32GB Sandisk SDCs: (i) a UHS (Ultra High Speed) class 1, and (ii) a UHS class 3.
Note that UHS 1 and UHS 3 refer to minimum write speed of 10 MB/sec and 30 MB/sec, respectively.
In addition to SDC speed, we are interested in measuring the effect of SDC capacity utilization on boot up performance.
Modern SDCs are implemented with the NAND technology.
Compared to the NOR technology, NAND offers lower power consumption, lower cost per bit, higher density and faster write speed.
However, as the disk fills up, the write performance starts to degrade.
In order to measure the effect of SDC utilization on boot up performance, we fill 5\%, 50\%, and 95\% of the SDC capacity.
To fill the SDC with data, we used the \texttt{dd} \cite{DDTool} utility which allows us to write blocks of memory to the SDC.

Tables \ref{tab:phases_rpi3} and \ref{tab:phases_rpizw} demonstrate the results averaged over 50 trials for each configuration.
The following unit configurations are used for these measurements: (i) EU, and (ii) ALLU w/o NET3 w/o WiFi).
We excluded NET3 due to the high variations caused by these services.
In terms of SDC capacity usage, these results show no effect on boot up performance.
However, the use of a faster SDC results in a slight reduction in energy consumption.
For example, for the ALLU w/o NET3 configuration on the RPi3, on average the faster SDC reduces energy by 2.5\%, compared to the slow SDC.

According to these results, regardless of the SDC type and capacity usage, $T_{btl}$ and $T_{knl}$ are similar for the two boards.
However, the energy consumption of these phases is higher on the RPi3 than on the RPiZW.
Compared to $P_{usi}$, which is process-intensive, $P_{btl}$ and $P_{knl}$ do not benefit from the higher processing power of RPi3's SoC because their main operation is to load the kernel and initialize hardware components. 
These are tasks that are mostly synchronous and not easily threaded across cores.
Therefore, since the RPi3 has a more complex and powerful SoC, more resources are wasted on this board during the first two phases.
In contrast, $E_{usi}$ of the RPi3 is actually lower than that of the RPiZW.
During $P_{usi}$, the RPi3 runs a larger number of processes in parallel, and therefore requires less energy to reach the boot up target by reducing the total amount of time spent in this phase.

\begin{table*}[]
\begin{center}
\definecolor{LightCyan}{rgb}{0.88,1,1}
    \caption{The duration and energy consumption of the bootloader phase, kernel phase, and userspace initialization phase for RPi3. The left and right values in each cell show duration (second) and energy consumption (joule).}
    \label{tab:phases_rpi3}
\begin{tabular}{cccccccc}
\cline{3-8}
                                                                                                    & \multicolumn{1}{c|}{}     & \multicolumn{3}{c|}{\textbf{EU}}                                                                              & \multicolumn{3}{c|}{\textbf{ALLU w/o NET3 w/o WiFi}}                                                           \\ \cline{3-8} 
                                                                                                    & \multicolumn{1}{c|}{}     & \multicolumn{1}{c|}{ \makecell{Bootloader+Kernel \\ ($T_{btl} + T_{knl}$, $E_{btl} + E_{knl}$)}} & \multicolumn{1}{c|}{\makecell{Userspace \\ ($T_{usi}$, $E_{usi}$)}} & \multicolumn{1}{c|}{Total}      & \multicolumn{1}{c|}{\makecell{Bootloader+Kernel \\ ($T_{btl} + T_{knl}$, $E_{btl} + E_{knl}$)}} & \multicolumn{1}{c|}{\makecell{Userspace \\ ($T_{usi}$, $E_{usi}$)}} & \multicolumn{1}{c|}{Total}       \\ \hline
\multicolumn{1}{|c|}{\multirow{4}{*}{\textbf{\begin{tabular}[c]{@{}c@{}}Slow \\ SDC\end{tabular}}}} & \multicolumn{1}{c|}{5\%}  & \multicolumn{1}{c|}{6.5, 4.51}     & \multicolumn{1}{c|}{2.94, 5.77}          & \multicolumn{1}{c|}{9.44, 10.28} & \multicolumn{1}{c|}{6.5, 4.51}     & \multicolumn{1}{c|}{4.04, 7.72}          & \multicolumn{1}{c|}{10.54, 12.23} \\ \cline{2-8} 
\multicolumn{1}{|c|}{}                                                                              & \multicolumn{1}{c|}{50\%} & \multicolumn{1}{c|}{6.5, 4.51}     & \multicolumn{1}{c|}{3.23, 5.95}          & \multicolumn{1}{c|}{9.73, 10.46} & \multicolumn{1}{c|}{6.5, 4.51}     & \multicolumn{1}{c|}{3.72, 7.32}         & \multicolumn{1}{c|}{10.22, 11.74} \\ \cline{2-8} 
\multicolumn{1}{|c|}{}                                                                              & \multicolumn{1}{c|}{95\%} & \multicolumn{1}{c|}{6.5, 4.51}     & \multicolumn{1}{c|}{3.23, 5.89}          & \multicolumn{1}{c|}{9.73, 10.4}  & \multicolumn{1}{c|}{6.5, 4.51}     & \multicolumn{1}{c|}{3.72, 7.28}          & \multicolumn{1}{c|}{10.22, 11.79} \\ \cline{2-8} 
\multicolumn{1}{|c|}{}                                                                              & \multicolumn{1}{c|}{AVG}  & \multicolumn{1}{c|}{6.5, 4.51}     & \multicolumn{1}{c|}{3.13, 5.87}          & \multicolumn{1}{c|}{9.63, 10.38} & \multicolumn{1}{c|}{6.5, 4.51}     & \multicolumn{1}{c|}{3.83, 8.91}          & \multicolumn{1}{c|}{10.33, 11.92} \\ \hline
                                                                                                    &                           &                                   &                                         &                                 &                                   &                                         &                                  \\ \hline
\multicolumn{1}{|c|}{\multirow{4}{*}{\textbf{\begin{tabular}[c]{@{}c@{}}Fast \\ SDC\end{tabular}}}} & \multicolumn{1}{c|}{5\%}  & \multicolumn{1}{c|}{6.5, 4.51}     & \multicolumn{1}{c|}{2.94, 5.6}           & \multicolumn{1}{c|}{9.44, 10.11} & \multicolumn{1}{c|}{6.5, 4.51}     & \multicolumn{1}{c|}{3.74, 7.13}          & \multicolumn{1}{c|}{10.24, 11.64} \\ \cline{2-8} 
\multicolumn{1}{|c|}{}                                                                              & \multicolumn{1}{c|}{50\%} & \multicolumn{1}{c|}{6.5, 4.51}     & \multicolumn{1}{c|}{2.97, 5.68}          & \multicolumn{1}{c|}{9.46, 10.19} & \multicolumn{1}{c|}{6.5, 4.51}     & \multicolumn{1}{c|}{3.66, 7.07}          & \multicolumn{1}{c|}{10.16, 11.58} \\ \cline{2-8} 
\multicolumn{1}{|c|}{}                                                                              & \multicolumn{1}{c|}{95\%} & \multicolumn{1}{c|}{6.5, 4.51}     & \multicolumn{1}{c|}{2.96, 5.67}          & \multicolumn{1}{c|}{9.46, 10.18} & \multicolumn{1}{c|}{6.5, 4.51}     & \multicolumn{1}{c|}{3.71, 7.2}           & \multicolumn{1}{c|}{10.21, 11.71} \\ \cline{2-8} 
\multicolumn{1}{|c|}{}                                                                              & \multicolumn{1}{c|}{AVG}  & \multicolumn{1}{c|}{6.5, 4.51}     & \multicolumn{1}{c|}{2.95, 5.65}          & \multicolumn{1}{c|}{9.45, 10.16} & \multicolumn{1}{c|}{6.5, 4.51}     & \multicolumn{1}{c|}{3.7, 7.133}          & \multicolumn{1}{c|}{10.2, 11.64}  \\ \hline
\end{tabular}
\end{center}
\end{table*}

\begin{table*}[]
\begin{center}
     \caption{The duration and energy consumption of the bootloader phase, kernel phase, and userspace initialization phase for RPiZW. The left and right values in each cell show duration (second) and energy consumption (joule).}
     \label{tab:phases_rpizw}
\begin{tabular}{cccccccc}
\cline{3-8}
                                                                                                    & \multicolumn{1}{c|}{}     & \multicolumn{3}{c|}{\textbf{EU}}                                                                               & \multicolumn{3}{c|}{\textbf{ALLU w/o NET3 w/o WiFi}}                                                           \\ \cline{3-8} 
                                                                                                    & \multicolumn{1}{c|}{}     & \multicolumn{1}{c|}{\makecell{Bootloader+Kernel \\ ($T_{btl} + T_{knl}$, $E_{btl} + E_{knl}$)}} & \multicolumn{1}{c|}{\makecell{Userspace \\ ($T_{usi}$, $E_{usi}$)}} & \multicolumn{1}{c|}{Total}       & \multicolumn{1}{c|}{\makecell{Bootloader+Kernel \\ ($T_{btl} + T_{knl}$, $E_{btl} + E_{knl}$)}} & \multicolumn{1}{c|}{\makecell{Userspace \\ ($T_{usi}$, $E_{usi}$)}} & \multicolumn{1}{c|}{Total}       \\ \hline
\multicolumn{1}{|c|}{\multirow{4}{*}{\textbf{\begin{tabular}[c]{@{}c@{}}Slow \\ SDC\end{tabular}}}} & \multicolumn{1}{c|}{5\%}  & \multicolumn{1}{c|}{6.5, 3.26}     & \multicolumn{1}{c|}{9.73, 7.02}          & \multicolumn{1}{c|}{16.23, 10.28} & \multicolumn{1}{c|}{6.5, 3.26}     & \multicolumn{1}{c|}{15.48, 10.91}        & \multicolumn{1}{c|}{21.88, 14.17} \\ \cline{2-8} 
\multicolumn{1}{|c|}{}                                                                              & \multicolumn{1}{c|}{50\%} & \multicolumn{1}{c|}{6.5, 3.26}     & \multicolumn{1}{c|}{9.43, 6.34}          & \multicolumn{1}{c|}{15.93, 9.6}   & \multicolumn{1}{c|}{6.5, 3.26}     & \multicolumn{1}{c|}{15.4, 10.97}         & \multicolumn{1}{c|}{21.9, 14.23}  \\ \cline{2-8} 
\multicolumn{1}{|c|}{}                                                                              & \multicolumn{1}{c|}{95\%} & \multicolumn{1}{c|}{6.5, 3.26}     & \multicolumn{1}{c|}{9.8, 6.57}           & \multicolumn{1}{c|}{16.3, 9.83}   & \multicolumn{1}{c|}{6.5, 3.26}     & \multicolumn{1}{c|}{15.47, 10.94}        & \multicolumn{1}{c|}{21.97, 14.2}  \\ \cline{2-8} 
\multicolumn{1}{|c|}{}                                                                              & \multicolumn{1}{c|}{AVG}  & \multicolumn{1}{c|}{6.5, 3.26}     & \multicolumn{1}{c|}{9.65, 6.64}          & \multicolumn{1}{c|}{16.15, 9.9}   & \multicolumn{1}{c|}{6.5, 3.26}     & \multicolumn{1}{c|}{15.45, 10.94}        & \multicolumn{1}{c|}{21.92, 14.2}  \\ \hline
                                                                                                    &                           &                                   &                                         &                                  &                                   &                                         &                                  \\ \hline
\multicolumn{1}{|c|}{\multirow{4}{*}{\textbf{\begin{tabular}[c]{@{}c@{}}Fast \\ SDC\end{tabular}}}} & \multicolumn{1}{c|}{5\%}  & \multicolumn{1}{c|}{6.5, 3.26}     & \multicolumn{1}{c|}{9.47, 6.56}          & \multicolumn{1}{c|}{15.97, 9.82}  & \multicolumn{1}{c|}{6.5, 3.26}     & \multicolumn{1}{c|}{15.3, 10.86}         & \multicolumn{1}{c|}{21.8, 14.12}  \\ \cline{2-8} 
\multicolumn{1}{|c|}{}                                                                              & \multicolumn{1}{c|}{50\%} & \multicolumn{1}{c|}{6.5, 3.26}     & \multicolumn{1}{c|}{9.41, 6.31}          & \multicolumn{1}{c|}{15.91, 9.57}  & \multicolumn{1}{c|}{6.5, 3.26}     & \multicolumn{1}{c|}{15.39, 10.92}        & \multicolumn{1}{c|}{21.89, 14.18} \\ \cline{2-8} 
\multicolumn{1}{|c|}{}                                                                              & \multicolumn{1}{c|}{95\%} & \multicolumn{1}{c|}{6.5, 3.26}     & \multicolumn{1}{c|}{9.47, 6.35}          & \multicolumn{1}{c|}{15.97, 9.61}  & \multicolumn{1}{c|}{6.5, 3.26}     & \multicolumn{1}{c|}{15.34, 10.92}        & \multicolumn{1}{c|}{21.84, 14.18} \\ \cline{2-8} 
\multicolumn{1}{|c|}{}                                                                              & \multicolumn{1}{c|}{AVG}  & \multicolumn{1}{c|}{6.5, 3.26}     & \multicolumn{1}{c|}{9.45, 6.41}          & \multicolumn{1}{c|}{15.95, 9.67}  & \multicolumn{1}{c|}{6.5, 3.26}     & \multicolumn{1}{c|}{15.34, 10.9}         & \multicolumn{1}{c|}{21.84, 14.16} \\ \hline

\end{tabular}
\end{center}
\end{table*}


\section{Profiling and Enhancement of the Linux Shutdown Phase}
\label{section:sys_shutdown}

Throughout the scope of this paper, special attention is paid to system boot up phases rather than the shutdown phase. 
This is because the time and energy required for boot up are higher than that required for the shutdown, and the gains are therefore more significant. 
However, there are also several ways through which shutdown time and energy consumption can be reduced. 
The naive approach is to cut the power to the RPi as soon as the user application is completed.
Unfortunately, cutting the power improperly might result in corrupting data blocks on SDC. 
If these blocks also happen to coincide with the sectors necessary for boot up or important files in the \texttt{rootfs}, the device could be rendered unrecoverable.
In order to address this problem, all of the SDC's partitions can be mounted as read-only to guarantee there would be no file operations when a power loss event occurs.
However, a read-only solution is complicated and not feasible when the user application needs to store or analyze large quantities of dynamic data such as running a machine learning algorithm.
Furthermore, since the filesystems are by default read-only, updating the device becomes a lengthy and challenging process, requiring virtual root filesystems mounted to RAM disks and multiple remount operations on the SDC. 
In this case, if large amounts of data must be stored for processing or before transmission, external storage is required.
This solution, however, introduces extra power consumption and might cancel out any gains achieved by a faster shutdown. 
Additionally, if no external device is used, workarounds must be implemented for system logging and other system functionalities, which rely on a writeable filesystem. 
This analysis is outside the scope of this paper.

The next approach is to make only the boot partition write-protected.
However, this solution does not necessarily prevent data corruption, because flash partitions are not actually separated as they would be on a traditional hard drive.
SDCs use a Flash Transition Layer (FTL) to map virtual file blocks to their actual location in the storage \cite{gupta2009dftl}. 
Many SDCs are preloaded with a wear-leveling firmware which uses the FTL to re-arrange data blocks, often mixing across partition lines (transparently to the RPi) to enhance block device lifetime.
In this case, data corruption can occur if the power is cut while read-only data is being migrated during the wear-leveling operation.
Therefore, mixing read-only and writeable partitions does not guarantee protection against improper shutdowns.
The only way to guarantee it is safe to cut the device power is to ensure that no operations are being performed on the SDC.

In the rest of this section, we present and evaluate two suitable approaches, \textit{graceful shutdown} and \textit{forced shutdown}, to power off a duty-cycled IoT system.
Furthermore, we assess the tradeoffs between system reliability and energy consumption.
Please note that $T_{sdn}$ and $E_{sdn}$ refer to the duration and energy consumption of the shutdown phase.


\subsection{Graceful Shutdown}
\begin{figure*}[t]
\centering
    \includegraphics[width=0.9\linewidth]{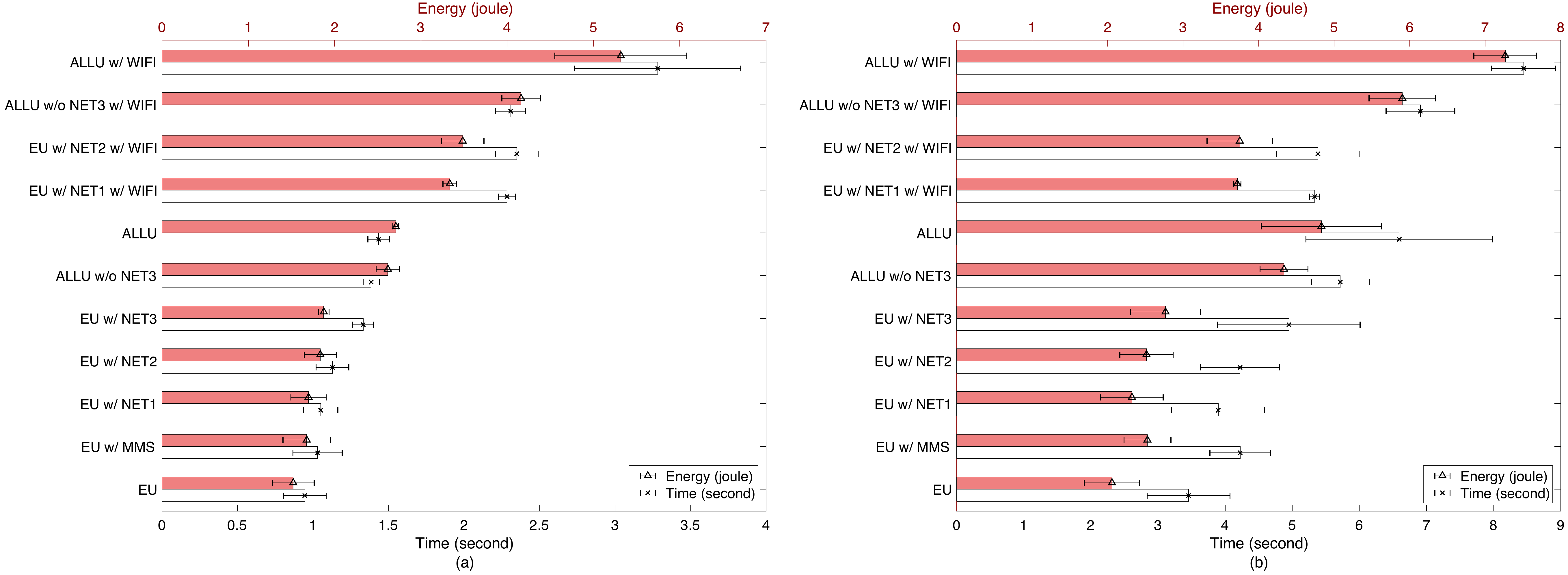}
    \caption{The duration and energy consumption of \textit{graceful shutdown} for the (a) RPi3 and (b) RPiZW when various unit configurations are applied. 
    The lower and upper axes show $T_{sdn}$ and $E_{sdn}$, respectively.}
    \label{fig:time_energy_three_zero}
\end{figure*}

During a graceful shutdown, \texttt{systemd} sends a shutdown signal to all of the running processes. 
After these processes exit and the network interfaces are brought down, the filesystems are unmounted and power to the device is safely cut. 
The amount of time required to unmount the filesystems is almost fixed and beyond the control of the user. 
However, the lower the number of running processes which must return an exit code before \texttt{shutdown.target} is reached, the faster \texttt{systemd} can finalize the shutdown; therefore, removing extraneous units expedites the shutdown phase.
This behavior is best exemplified in Figure \ref{fig:time_energy_three_zero}. 
These figures represent the time and energy consumption of the RPi3 and RPiZW for various unit configurations.
%

When WiFi is disconnected, using unit configuration EU reduces energy consumption by 43.9\% and 57.4\%, on the RPi3 and RPiZW, respectively, compared to unit configuration ALLU.
When WiFi is connected, using unit configuration EU reduces energy consumption by 37.3\% and 48.85\% for the RPi3 and RPiZW, respectively, compared to ALLU.
These results also show the significant effect of WiFi communication on $E_{sdn}$.
During the shutdown phase, the system invokes \texttt{ifdown} to ensure all connected networks are brought down properly and then powered off.
This process takes around 0.8s when \texttt{avahi-daemon.service} is disabled.
When this service is enabled, the duration varies depending on the network speed and configuration.
In our testbed, we noticed a delay of up to 12s.
For example, unit configuration EU w/ NET1 w/ WiFi increases energy consumption by 96.4\% and 60\%, for the RPi3 and RPiZW, respectively, compared to EU w/ NET1.




\subsection{Forced Shutdown}

Not all IoT applications require a clean shutdown to remain functional.
There are two alternative commands, \texttt{systemctl halt ----force} (equivalent to the traditional \texttt{halt} command) and \texttt{systemctl halt ----force ----force} (equivalent to the traditional \texttt{halt -f} command), which result in shorter shutdown phases.
In this paper we refer to these approaches as \textit{forced shutdown} and \textit{forced-forced shutdown}.
Both of these approaches skip the steps performed by \texttt{shutdown} to notify the running processes of the impending shutdown and wait for them to exit gracefully.
Therefore, steps such as recording the shutdown event and any \texttt{STDOUT} or \texttt{STDERR} output that are typically printed to a log file during the shutdown phase are skipped by these commands.
In the case of \texttt{systemctl halt ----force}, this may be acceptable, as the logging of system shutdown events is not often critical to IoT applications, and any necessary shutdown logs may be generated manually by the user application.
\texttt{systemctl halt ----force} also properly disconnects the networking interfaces by calling \texttt{ifdown} on all the connected interfaces.

\begin{figure}[!t]
\centering
    \includegraphics[width=1\linewidth]{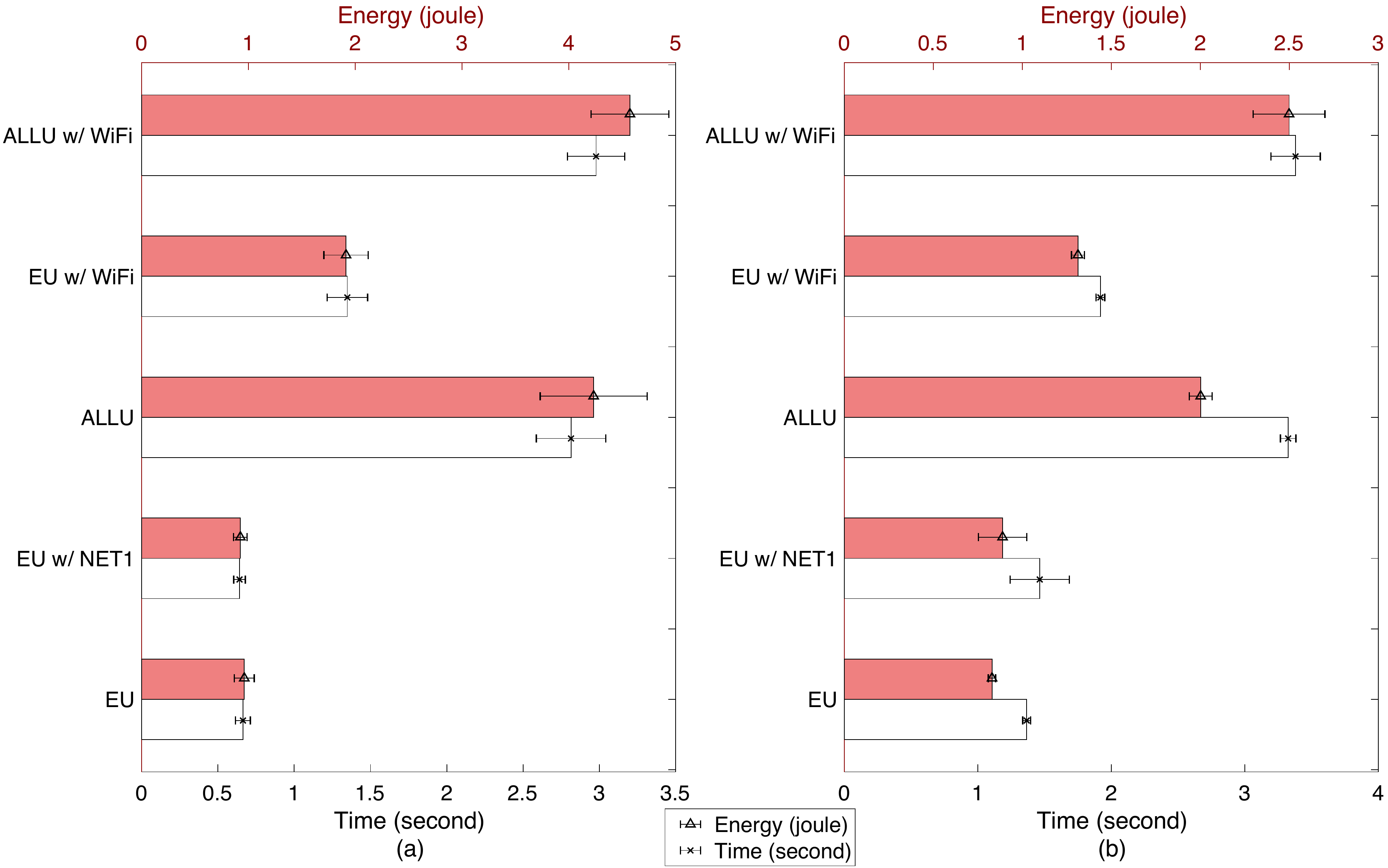}
        \caption{The duration and energy consumption of \textit{forced shutdown} for the (a) RPi3 and (b) RPiZW when various unit configurations are applied. 
    The lower and upper axes show $T_{sdn}$ and $E_{sdn}$, respectively.}
    \label{fig:shutdown_force}
\end{figure}

Figure \ref{fig:shutdown_force} shows the impact of using forced shutdown on both time and energy consumption.
Comparing this figure against Figure \ref{fig:time_energy_three_zero} demonstrates the performance benefits of killing processes rather than gracefully terminating them.
Although this mechanism results in lower energy consumption for both platforms, the effects are more apparent on the RPiZW.
When gracefully shutting down, the RPi3 provides the system processes with more resources to finalize their operations and terminate correctly, resulting in a shorter shutdown phase.
On the other hand, when forced shutdown is used, running processes are merely killed, which does not require significant system resources.
However, the OS still waits for the networking interfaces to be brought down before continuing the shutdown phase.
As a result, the extra power provided by the RPi3's processor is wasted, making the RPiZW more energy efficient during the shutdown phase.

\begin{figure}[!t]
\centering
    \includegraphics[width=1\linewidth]{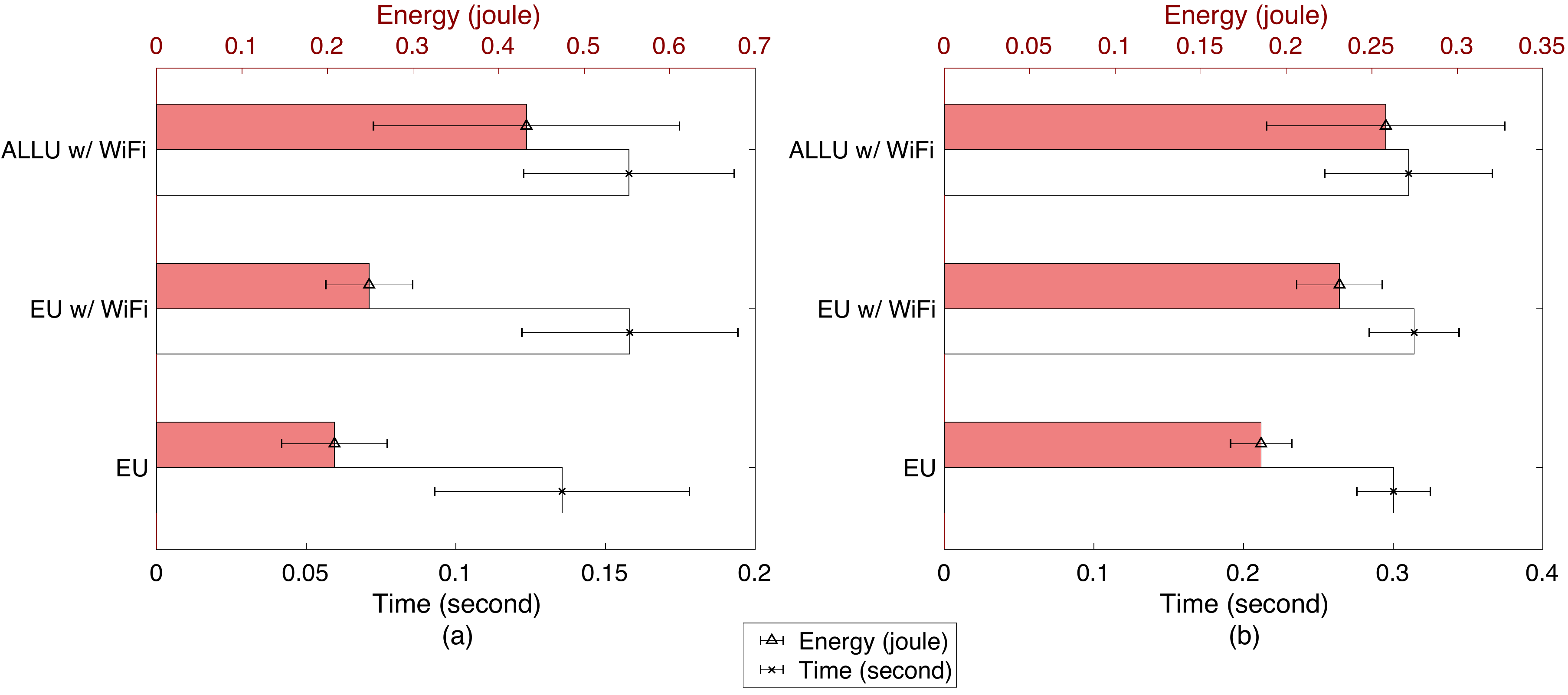}
        \caption{The duration and energy consumption of \textit{forced-forced shutdown} for the (a) RPi3 and (b) RPiZW when various unit configurations are applied. 
    The lower and upper axes show $T_{sdn}$ and $E_{sdn}$, respectively.}
    \label{fig:shutdown_force_force}
\end{figure}

The second approach, \texttt{systemctl halt ----force ----force}, is faster than calling \texttt{systemctl halt ----force} because the processes are not killed.
Instead, the processes are simply abandoned as the processor cores are stopped.
This command physically halts the processor and cuts power almost immediately without bringing down network interfaces or unmounting filesystems.
Eliminating these steps reduces the shutdown duration of the RPi3 and RPiZW to less than 200ms and 400ms, respectively, and the energy consumption to less than 700mJ and 350mJ, respectively, as Figure \ref{fig:shutdown_force_force} shows.
Unfortunately, according to the documentation \cite{systemctlService}, using this command may cause data corruption.
However, some approaches exist to minimize the risk.
For example, all processes required or started by the user application can be killed manually first, and the \texttt{sync} command must be run to commit unsaved buffers to the SDC.
For preventing issues related to wear-leveling, increasing the percentage of unused storage on SDC reduces the frequency of moving critical data blocks by the wear-leveling algorithm.
However, since the chance of data corruption is not fully eliminated, it is up to the user to calculate the risk involved in shutting down the system using this mechanism repeatedly (or across many devices). 
These calculations are outside the scope of this paper.
Additionally, for both forced-shutdown approaches, the user must consider running some processes (such as \texttt{fake-hwclock save} and \texttt{systemd-random-seed save}) manually to ensure system integrity and security.

It is worth mentioning that the \texttt{halt} system call for the ARM architecture automatically calls \texttt{machine\_power\_off()} to power off the board rather than entering the traditional \textit{halted state} where the board stays powered on after the processor is powered off.
Other architectures may require additional flags or even different commands in order to achieve a similar result.








\section{Parallelizing Application Processes with Userspace Initialization}

\label{section:PEF}

As we showed in the previous sections, the userspace initialization phase does not fully utilize the system resources.
In this section, we propose \textit{Pallex}, a parallel execution framework to run user applications during the userspace initialization phase.
After presenting the implementation of Pallex and providing guidelines for applying this framework, we evaluate its performance considering various IoT application scenarios.

\subsection{Pallex}
\label{pef}
The basic idea of Pallex is to divide a user application into \textit{stages} and run each stage based on the set of available units and the completion of prerequisite stages.
For a given user application, we break the code into a stage set $\mathrm{S} = \{s_{i}, s_{j},... \}$.
Each stage $s_{i}$ has two types of dependencies:
\begin{equation}
    \mathrm{D}(s_{i}) = \{s_{j}, s_{k}, ...\}    
\end{equation}
and
\begin{equation}
    \mathrm{D^{\prime}}(s_{i}) = \{u_{j}, u_{k}, ...\}
\end{equation}
where $\mathrm{D^{\prime}}(s_{i})$ is called the \textit{stage dependency set} and refers to the set of user application's stages that must be completed before starting stage $s_{i}$, and $\mathrm{D}(s_{i})$ is called \textit{unit dependency set} and refers to the set of units on which stage $s_{i}$ depends on. 
Each stage is invoked by \texttt{systemd} when the dependency sets specified in the stage's unit file are resolved.
Therefore, for \texttt{systemd} to build the dependency tree and run the stages in an orderly manner, each stage's dependency sets must be specified in its corresponding unit file's \texttt{Requires} section.

Since the stages of execution might be implemented as independent processes, they may not share the same address space.
Therefore, a mechanism is required to share data across processes.
To this end, Pallex utilizes the mechanism offered by \textit{Unix Domain Socket} (UDS) \cite{unixDomainSocket}.
UDS is a data communication endpoint for exchanging data between processes executing on the same host OS, and provides a standard method for implementing Inter-Process Communication (IPC) on a Unix-based system.
Since UDS handles communication within the Linux kernel, it is initialized in the kernel space as well.
This indicates that we can use them as reliable means of communication during $P_{usi}$.
When a stage $s_{i}$ completes its execution and needs to transfer data to another stage $s_{j}$, the kernel blocks $s_{i}$ until $s_{j}$ is ready \cite{bovet2005understanding}.
During this time, processor resources are used for the activation of units that are required by stage $s_{j}$.
When the dependencies of $s_{j}$ are resolved and it requests the shared data, $s_{i}$ completes the data transfer and then exits.


There are at least three other IPC methods we could have used, including TCP/UDP sockets, POSIX message queues, and writing to a file.
The reasons that we did not use these methods are as follows.
First, using a TCP/UDP connection over localhost requires waiting for \texttt{networking.service} to load, which reduces the level of concurrency for those stages that do not need the networking capability. 
Besides, assume that a user application only uploads the results to the cloud when abnormal behavior is detected. 
In this case, relying on the \texttt{networking.service} unnecessarily increases the energy consumption of the boot up process as we demonstrated in Section \ref{customize_units}).
Second, writing to a file increases the I/O overhead and affects the boot up time and message sharing delay because of the SDC activity during the userspace initialization phase.
Lastly, we did not use POSIX message queues because they are very low-level and require careful configuration.
Although both UDS and POSIX message queues are available almost concurrently, for the latter, it is necessary to configure the size and number of messages and their accepted/waiting status to make sure that enough buffer is available. 
Although message queues offer various features, UDS is standard, easy to use, and fast.
UDS allow us to send data to the socket before the receiving program is even started with little to no configuration. 
Therefore, the sending program can be completely divorced from the receiver regarding dependencies.
Also, UDS is agnostic towards payload size and can pause the sending program until the FIFO queue of data has begun to move and there is memory available to continue sending.

\begin{figure*}[!t]
\centering
    \includegraphics[width=0.8\linewidth]{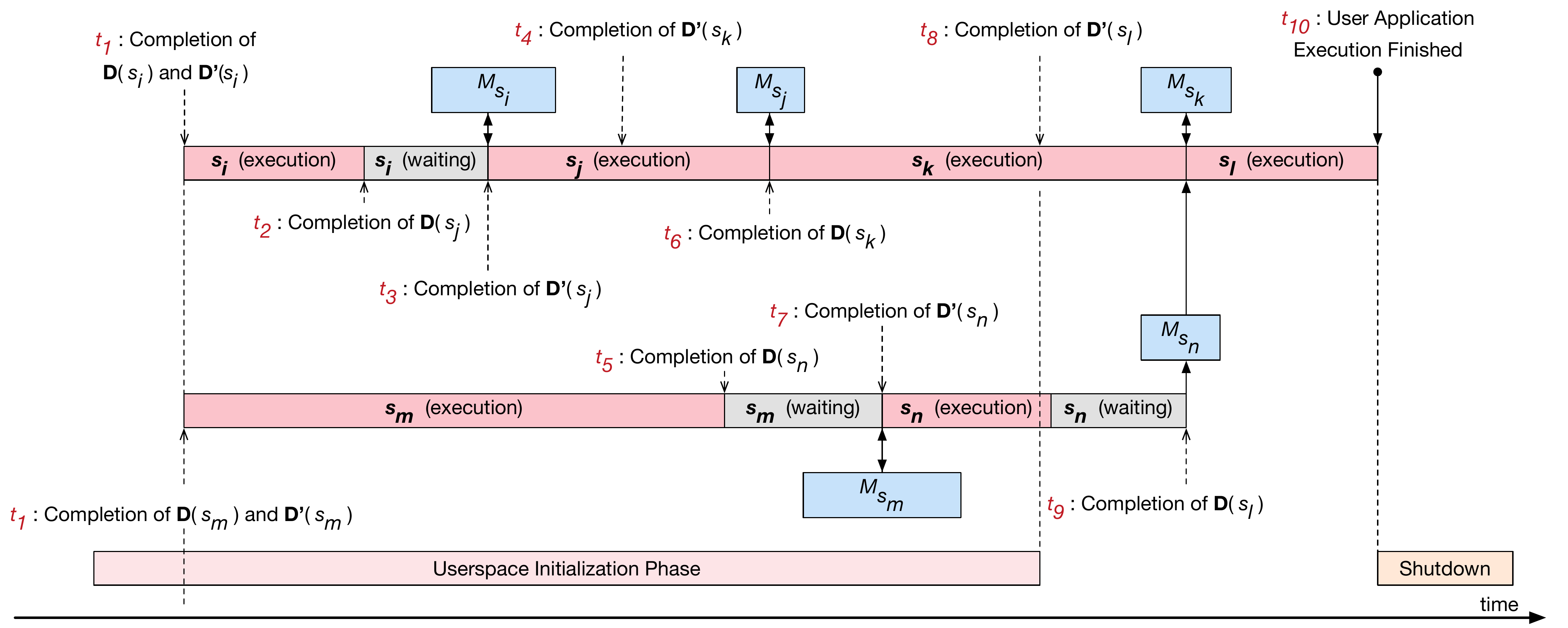}
    \caption{An example of Pallex. The \textit{stage set} of user application includes six stages, $\mathrm{S} = \{s_{i}, s_{j}, s_{k}, s_{l}, s_{m}, s_{n} \}$. A new stage is started as soon as its \textit{stage dependency set} and \textit{unit dependency set} are satisfied. Stages share their messages using the mechanism offered by Unix Domain Socket (UDS).}
    \label{fig:pef}
\end{figure*}

We further clarify the operation of Pallex through the scenario presented in Figure \ref{fig:pef}.
This example assumes that the user application is composed of two processes (or threads), and the first process depends on the data generated by the second process to complete its task.
Also, each process can be broken into a set of sequentially running stages.
For the stages we assume that,
\begin{gather*}
\mathrm{D}(s_{i}) = \emptyset,\;\; \mathrm{D}(s_{j}) = \{s_{i}\},\;\; \mathrm{D}(s_{k}) = \{s_{i}, s_{j}\},\\
\mathrm{D}(s_{m}) = \emptyset,\;\; \mathrm{D}(s_{n}) = \{s_{m}\}, \;\;
    \mathrm{D}(s_{l}) = \{s_{i}, s_{j}, s_{k}, s_{m}, s_{n}\}\;\; 
\end{gather*}
Each stage also depends on a set of units.
Figure \ref{fig:pef} shows the instances where the stage dependency set and unit dependency set of each stage are resolved.
Since $\mathrm{D}(s_{i}) = \{\}$ and $\mathrm{D}(s_{m}) = \{\}$, both stages can start at time $t_{1}$ at which point their unit dependency sets are resolved.
Therefore, these two stages run in parallel while the userspace units are being activated.
At time $t_{2}$, stage $s_{i}$ completes. 
However, $s_{j}$ cannot be started because its unit dependency set is resolved at $t_{3}$.
Therefore, $s_{i}$ is blocked and waits until stage $s_{j}$ is started and is ready to receive the data generated by $s_{i}$.
Sharing the messages generated by a stage $s_{i}$ with the next stages is denoted as $M_{s_{i}}$.
This figure also shows that, since $\mathrm{D}(s_{l}) = \{s_{i}, s_{j}, s_{k}, s_{m}, s_{n}\}$, stage $s_{l}$ cannot be started before the completion of $s_{n}$.
Once started, stage $s_{l}$ reads $M_{s_{n}}$ and $M_{s_{k}}$, the messages shared by two stages $s_{n}$ and $s_{k}$.
At time $t_{9}$, both the unit and stage dependency sets of $s_{l}$ are resolved, and this stage has all the resources necessary to complete its operation.
The user application finishes at time $t_{10}$. 
At this point, the system enters the shutdown phase.
%

%
%
%

The performance improvement achieved by Pallex highly depends on the number of and dependencies between the stages of a user application.
Specifically, to minimize the waiting time of each stage and reduce energy consumption, it is important to break a user application into stages with small unit dependency sets.
To this end, system developers must identify the main tasks performed by the application carefully, before restructuring them as stages.
Each stage must resolve all its dependencies before beginning execution.
Therefore, each stage must comprise a set of instructions that once started, can be completed without relying on the completion of any other unit or stage.
However, finding the right decomposition might not be a straightforward task due to the large number of units activated during $P_{usi}$.
Besides, little to no improvement is observed by minimizing the waiting time of stages on units requiring execution time of only a few milliseconds.
Due to the sampling-processing-sending nature of IoT applications, we can narrow down the list of essential units to simplify the task of decomposition, as follows.


\begin{itemize}
\item [--] {GPIO.} These interfaces are initialized by the GPU. 
Therefore, as soon as \texttt{systemd} finishes its initialization (0.75s for the RPi3 and 1.2s for the RPiZW), user applications can use the GPIO pins.
In addition to enabling the execution of user applications at the beginning of $P_{usi}$, using GPIOs provides a faster communication interface without the high overhead of a full networking stack, especially when a small amount of data must be communicated between nearby devices.
\item [--] {I2C, SPI, CSI.} The drivers for I2C (Inter-Integrated Circuit), SPI (Serial Peripheral Interface) and CSI (Camera Serial Interface) are loaded as kernel modules during $P_{knl}$.
Therefore, the unit dependency of user applications relying on these components is resolved at the beginning of $P_{usi}$.
%
%
%
%
%
%
\item [--] {Bluetooth.} Bluetooth depends on both \texttt{hciuart.service} and \texttt{bluetooth.service}.
As Section \ref{section::performanceEnhancement} showed, the activation of these services finishes after \texttt{rc.local}.
Therefore, user application stages that rely on Bluetooth must be started by creating a \texttt{systemd} service that starts after \texttt{bluetooth.service} instead of \texttt{rc.local}.
If an application includes tasks that do not depend on these services, then running those tasks as stages that start before the completion of these services can result in a considerable performance improvement, especially due to their long activation duration.
\item [--] {WiFi.}
Stages that rely on WiFi must be initialized after the activation of \texttt{network.target}.
Another option is to start the stage after \texttt{network-online.target}, which is invoked once the network is \textit{connected} as opposed to \textit{available}. 
\item [--] {Ethernet.}  
Similar to WiFi, the status of the Ethernet interface can be derived from \texttt{network.target}.
A difference between the WiFi and Ethernet interfaces is that the speed of initializing Ethernet is faster because it does not perform the authentication and association process required for WiFi.
Therefore, using Ethernet results in a shorter duty cycle.
However, since most IoT applications rely on wireless communications, we mainly focus on WiFi in this paper.
%
%
%
%
\end{itemize}

In Appendix \ref{appendix:services}, we present an overview of units to provide the users with guidelines regarding the impact of each unit on each application scenario.

\subsection{User Application Scenarios for Evaluating Pallex}
\label{PEF_scenarios}
In this section, we evaluate the performance of Pallex when applied to various types of user applications implemented on the RPi3 and RPiZW.
Since using the RPi3 or RPiZW is justified when the application at hand cannot be accomplished using resource-constrained devices (such as those employing ARM Cortex-M or R processor), our user application scenarios include heavy operations like image capture, encryption, and classification.
However, it should be noted that we omit image classification using RPiZW due to the high overhead caused by running the machine learning algorithm on this platform.
We explain these scenarios in the following subsections.


\subsubsection{Scenario 1: Image Capture (IC)}  For this scenario we used a camera module \cite{RPiCam} to capture an image.
The application stage set includes only one stage, $\mathrm{P} = \{s_{cap}\}$, where $\mathrm{D}(s_{cap}) = \emptyset$ and $\mathrm{D^{\prime}}(s_{cap}) = \emptyset$.
Therefore, we use unit configuration EU for this scenario.
The camera module is capable of capturing one picture per second (each around 2.5MB) and uses CSI to communicate with the RPi. 
After capturing the image, the user application saves the image as a JPEG file.
The stage $s_{cap}$ is written as a Python program that calls the picamera library \cite{picamera} (written in Python) to capture the image.
\subsubsection{Scenario 2: Image Capture+Encryption (IC\&E)}
We extend "Scenario 1" by encrypting the captured image using the AES-256 encryption algorithm.
The application stage set includes one stage, $\mathrm{P} = \{s_{cap+enc}\}$, where $\mathrm{D}(s_{cap+enc}) = \emptyset$ and $\mathrm{D^{\prime}}(s_{cap+enc}) = \emptyset$.
Please note that since encryption depends on capture, we do not decompose the application into two stages. 
This scenario uses Unit configuration EU.
The stage $s_{cap+enc}$ is written as a Python program that calls the picamera library to capture the image, then calls the GNU Privacy Guard (\texttt{gpg}) utility (written in C) to encrypt the image.
\subsubsection{Scenario 3: Image Capture+Classification (IC\&C)}
In this scenario, we capture an image and classify it using a pre-trained K-Nearest Neighbors (KNN) algorithm \cite{altman1992introduction}.
Since image capture and loading the model are independent, we decompose the application into three stages: image capture ($s_{cap}$), loading the KNN model ($s_{load}$), and performing classification ($s_{clas}$).
Therefore, $\mathrm{P} = \{s_{cap}, s_{load}, s_{clas}\}$, where $\mathrm{D}(s_{cap}) = \emptyset$, $\mathrm{D^{\prime}}(s_{cap}) = \emptyset$, $\mathrm{D}(s_{load}) = \emptyset$, $\mathrm{D^{\prime}}(s_{load}) = \emptyset$, $\mathrm{D}(s_{clas}) = \{ s_{cap}, s_{load} \}$ and $\mathrm{D^{\prime}}(s_{cap}) = \emptyset$.
Please note that $s_{cap}$ and $s_{load}$ are two concurrently running processes.
Consequently, the two heavy stages, i.e., image capture and loading the model, start and run concurrently.
$s_{clas}$ depends on the completion of $s_{cap}$ and $s_{load}$  to perform its operation.
This scenario uses Unit configuration EU.
The implementation of the $s_{cap}$ stage is as before.
The stage $s_{load}$ is implemented in a Python program that loads scikit-learn's KNN model (written in Cython) \cite{scikit-learn}.
The stage $s_{clas}$ is implemented in a Python program that uses scikit-learn's KNN algorithm to classify the image.
\subsubsection{Scenario 4: Image Capture+Upload (IC\&U)}
In this scenario, we capture an image and transmit it to a cloud server through WiFi communication with an access point.
The application stage set includes two stages, $\mathrm{P} = \{s_{cap}, s_{upl}\}$, where $\mathrm{D}(s_{cap}) = \emptyset$, $\mathrm{D^{\prime}}(s_{cap}) = \emptyset$, $\mathrm{D}(s_{upl}) = \{ s_{cap} \}$ and $\mathrm{D^{\prime}}(s_{upl}) = \{ u_{net}\}$, where $u_{net}$ refers to the \texttt{network-online.target} unit.
Please note that the service configuration we used for this scenario is EU w/NET1.
%
Because networking services require a relatively long duration to initialize, we are interested in evaluating Pallex in scenarios where $T_{usi}$ is long.
For the RPi3 and RPiZW, we vary the number of images that are captured and uploaded from 1 to 3.
For the RPi3, activating \texttt{networking.service} is approximately 3.5s, which means that we can capture a maximum of 3 images in parallel with this service activation without negatively impacting the duration of userspace initialization.no
Although activating \texttt{networking.service} on the RPiZW is longer than 7.4s, we cap the maximum number of images at 3 to present a fair comparison across the two boards.
These sub-scenarios are referred to as IC\&U$x$, where $x$ refers to the number of images captured and uploaded.
The implementation of the $s_{cap}$ stage is as before.
The stage $s_{upl}$ is implemented as a Python program that calls the \texttt{curl} utility (written in C) to upload the image.

\begin{figure}[!t]
\centering
    \includegraphics[width=0.9\linewidth]{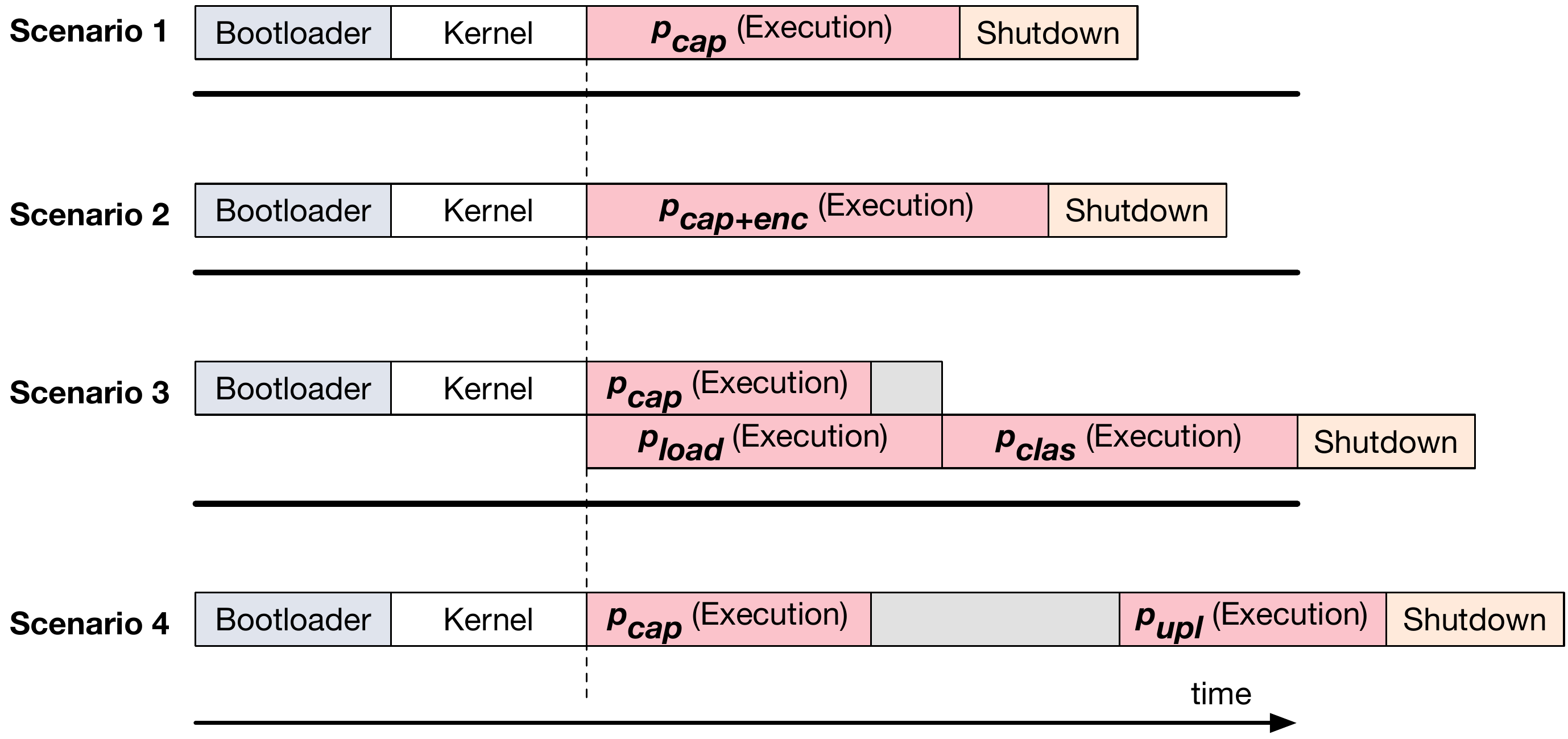}
    \caption{The four scenarios used to measure the effect of Pallex on duration and energy consumption. 
    We have used three variations of Scenario 4 to evaluate Pallex when using WiFi for uploading data.}
    \label{fig:pef_scenarios}
\end{figure}

%
Figure \ref{fig:pef_scenarios} shows a summary of the operations of these applications versus time.

\textbf{Threats to Validity.}
As the authors in \cite{georgiou2018your} showed, the choice of programming language affects the energy consumption of user application depending on the task being performed.
This must be taken into account when the measurements are repeated using different programming languages than those we used.

\subsection{Results}

Considering the user application scenarios given in Section \ref{PEF_scenarios}, we present the performance measurement results of Pallex in this section.
It must be noted that Pallex is compared against baseline scenarios that employ unit configuration to prevent the activation of unnecessary units.
In the baseline scenarios, referred to as "EU" in the figures, the EU configuration is used, and the user application is started when \texttt{rc.local} is loaded.

Regarding duration and energy measurement, we consider the interval between the instance the RPi is powered on until the completion of the user application program.
Please note that merely measuring the impact of Pallex on the duration of completing a user application does not reflect performance in terms of energy efficiency and lifetime.
Since the OS dynamically adjusts the operating frequency of the processor cores based on load \cite{brodowski2015cpu} (a.k.a., dynamic frequency scaling), it is critical to verify that the additional system load during $P_{usi}$ does not eliminate the energy saving achieved by reducing the duration.

\begin{figure}[!t]
\centering
    \includegraphics[width=1.0\linewidth]{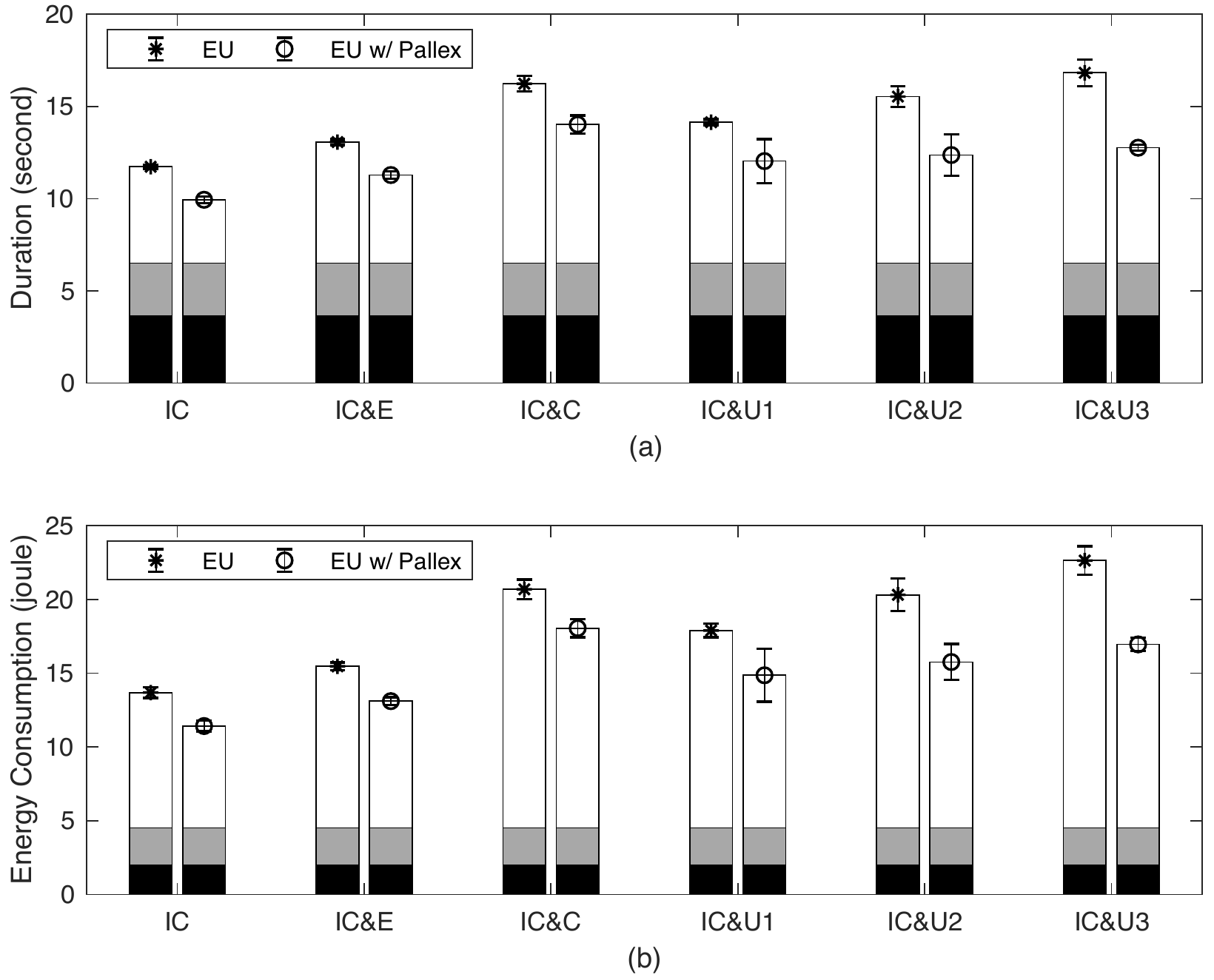}
    \caption{The duration and energy consumption of Pallex versus normal launching of user applications. 
    The device used is a RPi3.
    Black, grey, and white bars show the bootloader phase, kernel phase, and the completion of user application.}
    \label{fig:cef_three}
\end{figure}

\begin{figure}[!t]
\centering
    \includegraphics[width=1.0\linewidth]{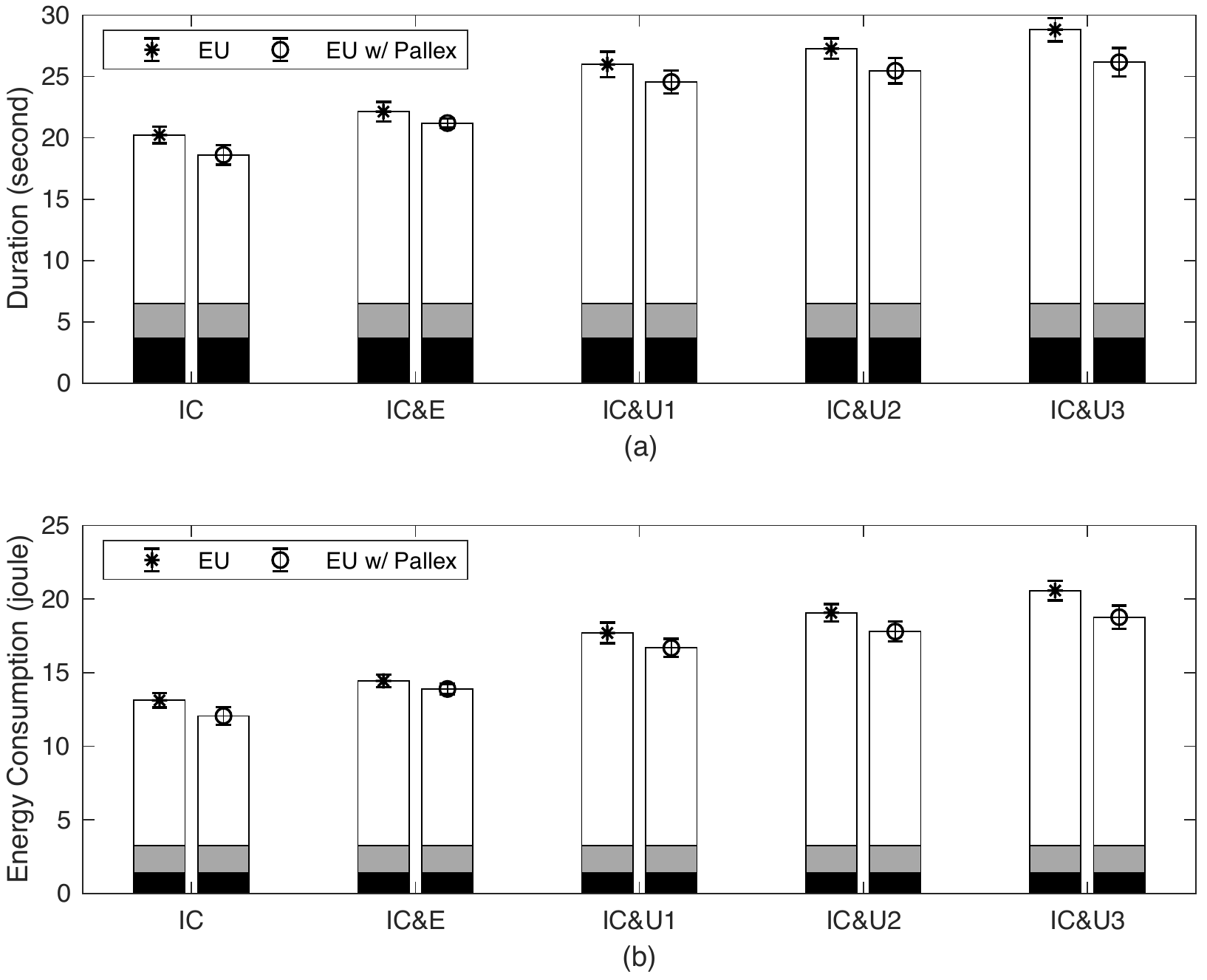}
    \caption{The duration and energy consumption of Pallex versus normal launching of user applications. 
    The device used is a RPiZW.
    Black, grey, and white bars show the bootloader phase, kernel phase, and the completion of user application.}
    \label{fig:cef_zero}
\end{figure}

Figure \ref{fig:cef_three} and \ref{fig:cef_zero} show the performance improvements achieved using Pallex for RPi3 and RPiZW during the boot up process, respectively. 
In these figures, the black and gray bars represent $P_{btl}$ and $P_{knl}$, respectively.
It must be noted that, since Pallex affects system performance after $P_{knl}$, both $E_{btl}$ and $E_{knl}$ are fixed irrespective to the RPi board used.
From the userspace processing point of view, for the IC scenario, energy consumption is reduced by 24.7\% and 10.77\% for the RPi3 and RPiZW, respectively.
For the IC\&E scenario, we observe 27.22\% and 5.98\% improvement in terms of duration and 21.54\% and 4.89\% in terms of energy for the RPi3 and RPiZW, respectively.
Both of these scenarios include a single stage that is executed as soon as \texttt{systemd} is ready.
It must be noted that the user application is not the only process running after the completion of \texttt{systemd} initialization.
As explained in Section \ref{customize_units}, the units that are essential for maintaining system integrity and stability are being activated during this duration as well.

Each IC\&C scenario is composed of three stages, where two stages run in parallel and during $P_{usi}$.
For this scenario we observe a 22.68\% improvement in terms of duration and 16.29\% in terms of energy for the RPi3.

The IC\&U scenario is composed of two stages, where the first stage runs during $P_{usi}$.
As we showed in the previous sections, activating the \texttt{networking.service} is a lengthy process due to the initialization of hardware and networking utilities as well as association with the access point.
Therefore, this scenario can significantly benefit from capturing images while the activation of networking services is in progress. 
In addition, more improvement is observed as the number of captured images increases:
When one image is captured, we find 27.57\% and 7.3\% improvement in terms of userspace processing duration for the RPi3 and RPiZW, respectively.
This results in an energy improvement of 22.67\% for the RPi3 and 7.04\% for the RPiZW.
These improvements are increased to 39.36\% and 11.89\% for these two boards when three images are captured.
Regarding energy improvement, we observe 31.35\% and 10.45\% improvement for the RPi3 and RPiZW, respectively, when three images are captured.

Comparing the two hardware platforms, we can observe that when Pallex is applied, the RPi3 shows a higher performance improvement compared to the RPiZW.
These results are consistent with our observations in Section \ref{customize_units}, indicating that the RPi3 platform consumes less energy than the RPiZW in a duty-cycling capacity despite drawing more current.
The RPi3 parallelizes userspace initialization processes across multiple cores, resulting in a shorter duty-cycle duration.
The impact of shortening the processing duration is more significant than the impact of the difference in current consumption across platforms.





\begin{figure}[!t]
\centering
    \includegraphics[width=1.0\linewidth]{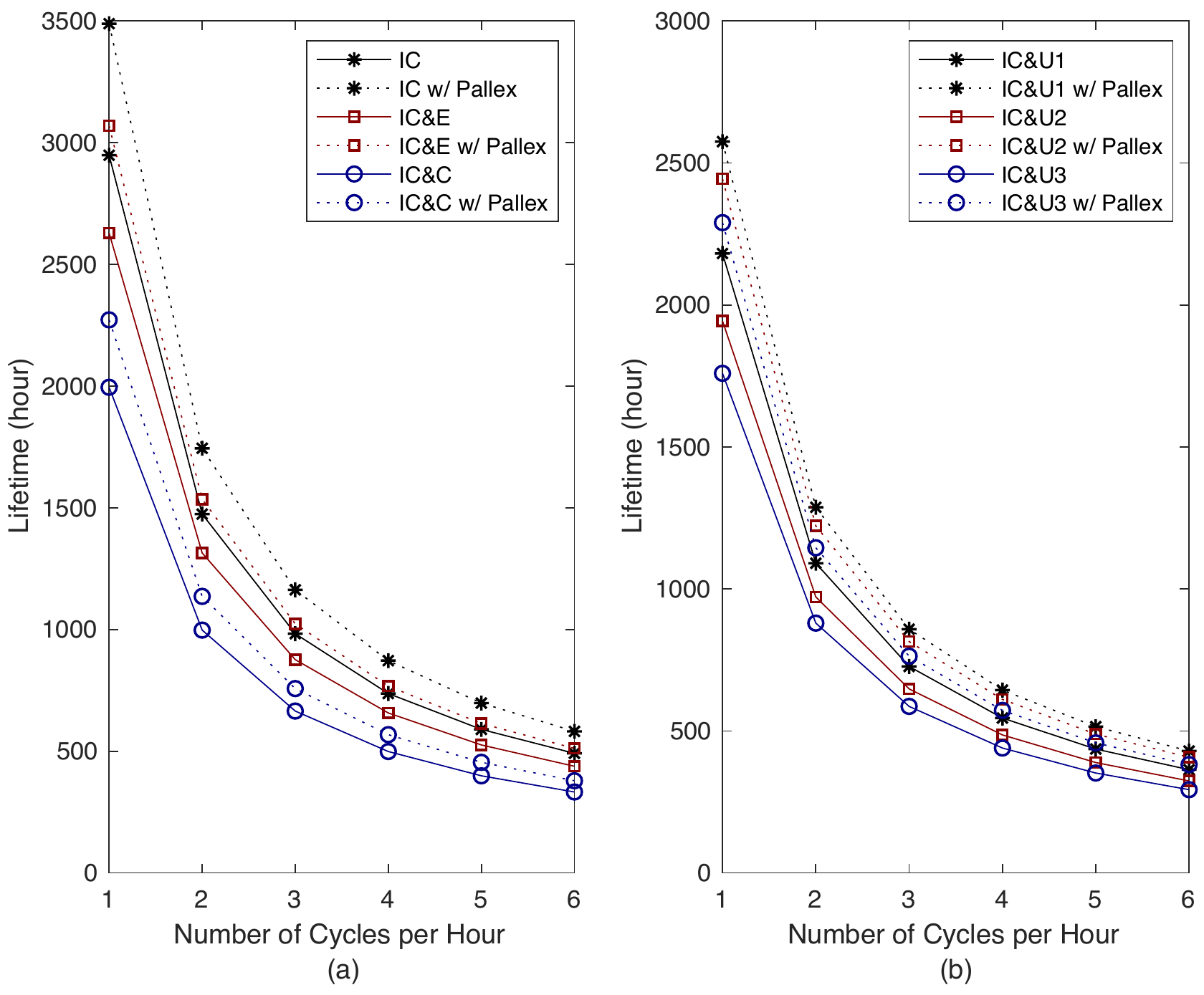}
    \caption{Lifetime of the RPi3 for different user application scenarios.
    Sub-figure (a) shows networking-independent scenarios, and sub-figure (b) shows networking-dependent scenarios.
    }
    \label{fig:lifetime_three}
\end{figure}

\begin{figure}[!t]
\centering
    \includegraphics[width=1.0\linewidth]{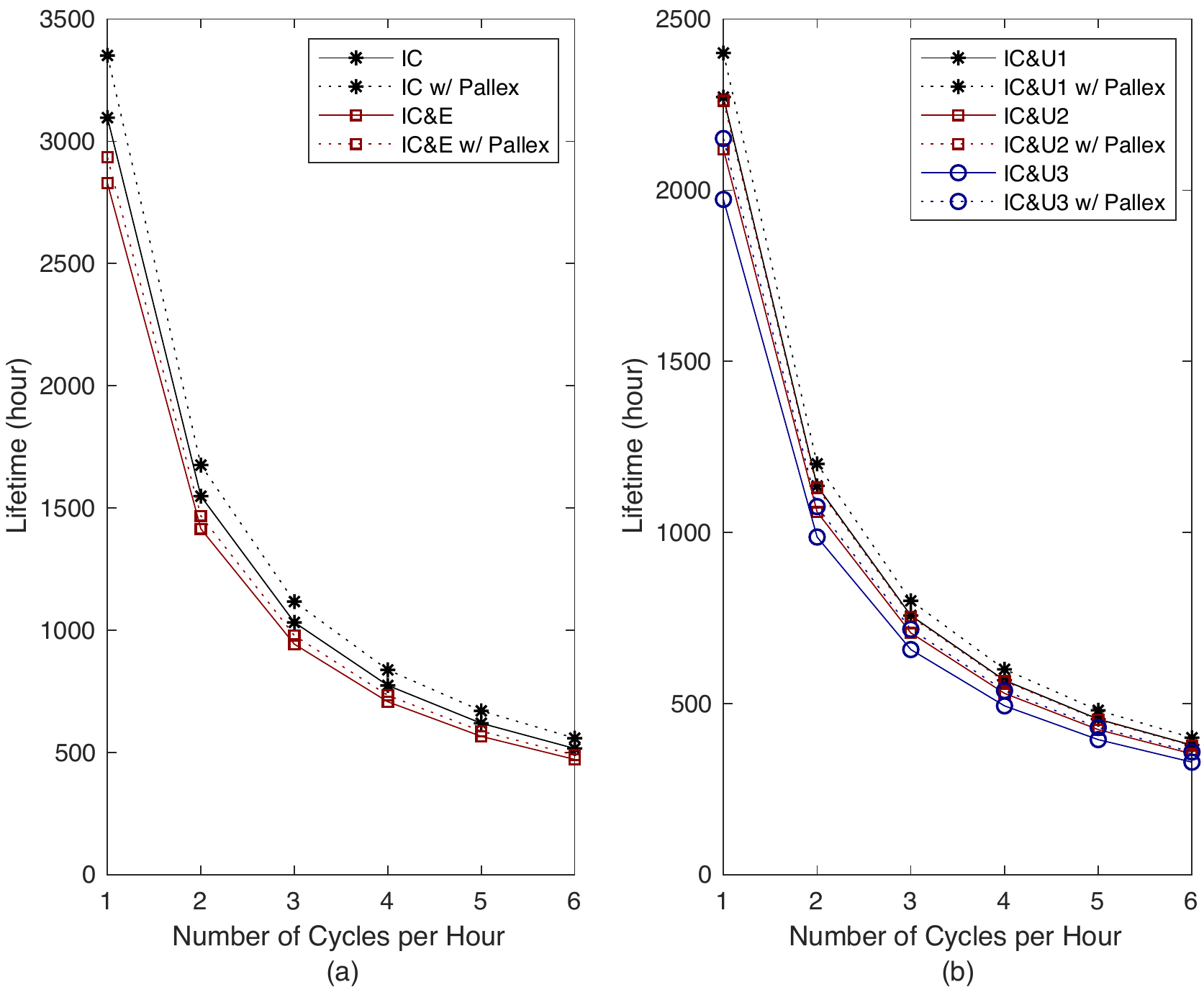}
    \caption{Lifetime of the RPiZW for different user application scenarios. 
    Sub-figure (a) shows networking-independent scenarios, and sub-figure (b) shows networking-dependent scenarios.
    }
    \label{fig:lifeftime_zero}
\end{figure}

To measure the impact of Pallex on the lifetime of duty-cycled systems, we compute system lifetime as follows:
\begin{equation}
\begin{split}
        \mathrm{lifetime} = \frac{E_{bat}}{(E_{btl} + E_{knl} + E_{usr} + E_{sdn}) \times N}\\ = \frac{3600 \times 2400 \times 10^{-3} \times 5}{(E_{btl} + E_{knl} + E_{user}+ E_{sdn}) \times N} 
\end{split}
\end{equation}
where $E_{btl}$, $E_{knl}$, $E_{user}$, and $E_{sdn}$ are the energy consumption of the bootloader phase, kernel phase, user application, and shutdown phase.
$E_{bat}$ is the available energy of the battery, and $N$ is the number of cycles per hour.
For the shutdown phase, we used the \textit{forced shutdown} mechanism detailed in Section \ref{section:sys_shutdown}, as it is faster than the \texttt{shutdown} command without sacrificing reliability.
We also assume the capacity of the battery is 2400mAh and its voltage is 5V.
Figures \ref{fig:lifetime_three} and \ref{fig:lifeftime_zero} show system lifetime versus the number of cycles per hour for the RPi3 and RPiZW, respectively.

The maximum improvement in the lifetime for the RPi3 is 30.16\% for scenario IC\&U3, and the minimum is 13.89\% for scenario IC\&C.
Similarly, for the RPiZW, the maximum improvement is 9.01\% for scenario IC\&U3, and the minimum is 3.74\% for scenario IC\&C.
For the networking-dependent scenarios, we notice that lifetime improvement increases versus the number of images captured and transmitted.
Specifically, for the RPi3, we notice an improvement of 18.06\% for IC\&U1, 25.74\% for IC\&U2, and 30.16\% for IC\&U3.
For the RPiZW, we notice an improvement of 5.67\% for IC\&U1, 6.7\% for IC\&U2, and 9.01\% for IC\&U3.
For the networking-independent scenarios, IC\%C achieves the highest increase in the lifetime (18.33\%) on the RPi3.
This phenomenon is attributed to the higher processing demand and concurrent execution of $s_{cap}$ and $s_{load}$ stages.

The improvements in the lifetime, in particular, reduce the cost of energy harvesting systems as well as system maintenance.
For example, when the amount of energy consumed per hour is reduced, a lower-cost energy harvesting system (e.g., smaller solar panels, smaller batteries) can be used as it is provided with more time to harvest and store energy.
For fully battery powered systems, increasing lifetime reduces the frequency of system maintenance to replace or recharge the batteries.





%

%








\section{Threats to Validity}
\label{threats_validity}
In addition to the discussions regarding threats to validity presented in the previous sections, we highlight additional internal and external threats in this section.

\textbf{Internal Validity.} The power measurement tool used for the experiments of this paper presents an accuracy of 2.5\%, as proved in \cite{dezfouli2018empiot}, when used in the room temperature of 23\textdegree{}C.
Nevertheless, it should be noted that both the accuracy of the power measurement tool as well as the energy consumption of the board vary with temperature.
Specifically, for the power measurement tools without a self-calibration feature, the impact of temperature on analog-to-digital converter (ADC) accuracy must be carefully taken into account.

\textbf{External Validity.} As mentioned in Sections \ref{section:method} and \ref{profiling_energy_time_btl_knl}, the experiments of this paper have been conducted using two different hardware platforms, one OS, and two different SDCs.
Future work could address the applicability of these observations using different hardware platforms and software environments.
Regarding OS, we have employed the March 2018 release of RSL.
However, it must be noted that the results might be different for the new releases of this OS.
A study similar to that performed in \cite{hindle2015green} can be very helpful to reveal the impact of OS maintenance and updates on performance.

\textbf{Reliability.} The testbed architecture and experimentation methodology have been presented in Section \ref{section:method} in details. 
This enables the extension of this work to other systems and scenarios while establishing a clear baseline for comparison with our work.

\section{Related Work}
\label{section::related}

In this section, we overview the existing works relevant to energy measurement and modeling, boot up improvement, and energy efficient software development.

\subsection{Energy Measurement and Modeling}
\label{related_energy_model}
Although most of the existing works on the energy measurement and modeling of low-power systems focus on the communication cost \cite{vilajosana2014realistic,wang2006realistic}, there exist a few studies that model the energy consumption of the whole system.
The authors in \cite{martinez2015power} present a systematic analysis of energy distribution in IoT devices that rely on energy harvesting.
Although the accuracy of the proposed models have been verified against empirical evaluations, the proposed models are only applicable to IoT devices that utilize micro-controllers (e.g., Cortex-M4) and simple OSs (e.g., RTOS).
In contrast, the RPi3 board includes a Cortex-A53 processor and the operation of its OS is significantly more complex than RTOSs such as FreeRTOS \cite{FreeRTOS} and ThreadX \cite{ThreadX}.
PowerPi \cite{kaup2014powerpi} proposes a power estimation model for the RPi Model B.
The power estimation model is composed of the power consumption of a processor, an Ethernet NIC, and an external WiFi transceiver.
Although the proposed models present a realistic power consumption of the board versus processor and communication load, the models do not provide insights regarding the performance of duty-cycled systems regarding the boot up overhead.
Besides, applying these models requires the pre-measurement of processor utilization and networking load, which might not be feasible for complex tasks.
An extension of the PowerPi modeling approach has been proposed in \cite{kaup2018progress} for various single-board computers.
The authors in \cite{ARMProfiling} model the power consumption of ARM-based architecture, using regression analysis to develop statistical power models. 
Attention is given to both the processor itself and the system as a whole, i.e. wireless radios, in different scenarios, such as browsing the web or multimedia benchmarking. 
The paper concludes that hardware changes such as increasing the size of the L2 cache can increase power efficiency, but does not go into detail concerning improvements that can be made in software to increase efficiency when modifying the hardware is not feasible.

\subsection{Boot-up Improvement}
In this section, in addition to userspace optimization, we also study several other techniques for improving boot up, namely, bootloader optimization, kernel space optimization, and suspend optimization.
Variants of these techniques can be utilized in conjunction with our work to further enhance boot up time and energy consumption.

\subsubsection{Userspace Optimization}
The authors in \cite{Villegas2006ImproveTD} identify several areas of improvement for boot up time on Debian, a Linux operating system, which at the time used \texttt{sysvinit}.
Although Raspbian uses \texttt{systemd} instead of \texttt{sysvinit}, its backwards compatibility allows for some of these optimizations to translate over to devices using \texttt{systemd} as their \texttt{init} system.
The first area of optimization involves substituting lighter default applications.
For example, boot up time was improved by two seconds by using \texttt{dash} instead of \texttt{bash}, the default shell for Debian.
Rewriting slow shell scripts to use internal functions instead of external functions has also been proven to improve the performance of boot up time.

Boot scripts can also be executed in parallel.
This is achieved through setting the \texttt{CONCURRENCY} option in each boot script.
Reordering boot scripts can also help programs complete faster.
In \cite{jo2009optimizing}, the authors identified areas during the boot up process of a smart TV when processor utilization is low.
They used this information to optimize boot script execution by interleaving I/O-intensive programs with processor-intensive programs.
While this can be an effective solution, maintaining system stability can be troublesome, as \texttt{sysvinit} does not track dependencies.
Therefore, it is the user's responsibility to ensure any changes do not compromise system stability.
Compounding the issue, dependencies between boot scripts are often not well documented and require thorough research and an understanding of both the Linux kernel and \texttt{sysvinit} before modification is possible.
By comparison, \texttt{systemd}, a modern \texttt{init} system, tracks dependencies and automatically generates a tree structure to determine which startup tasks can be executed in parallel.
Therefore, the feature set of \texttt{systemd} makes boot script reordering a much simpler task.

In addition to improving boot up time, we studied several methods for shortening the duration until \textit{user interaction time}: the time it takes until the user is able to interact with the application.
In \cite{Villegas2006ImproveTD} the authors are able to gain an improvement of two seconds by performing certain tasks, such as setting up the network and hardware clock, in the background.
In \cite{singh2011optimizing, jo2009optimizing}, the authors optimize boot up time on Android smartphones and televisions by deferring long-running services until after the log-in screen is displayed to the user.
However, this approach is not directly applicable to the IoT devices employing duty cycling.


\textit{Application XIP} (Application Execute in Place) \cite{Bird2010MethodsTI} is another technique for improving application start up speed.
\textit{Application XIP} allows for user applications to be executed directly from the filesystem instead of first being loaded into RAM.
When a program is executed, the kernel program loader maps text segments for applications directly, resulting in a reduction in RAM footprint, faster first invocation, and reduced power consumption of flash vs RAM.
However, \textit{Application XIP} requires compiling the kernel with support for a filesystem (such as CRAMFS \cite{cramfs}) that supports storing files in contiguous blocks of memory.
Additionally, there are hardware requirements for XIP such as random access storage that is directly accessible by the processor, which the RPi platform does not offer.

One major complexity of userspace initialization can be attributed to the evolution of the \texttt{init} system's responsibilities.
Modern \texttt{init} systems often manage process scheduling, I/O scheduling, and memory scheduling, among other responsibilities.
In \cite{Lim2015EnhancingIS}, the authors addressed this issue by simplifying the existing \texttt{init} scheme to improve an Android device's boot up time.
Their contribution involves separating the \texttt{init} scheme into two booting modes: normal boot, which executes all tasks during boot up time, and quick boot, which executes only mandatory tasks before user interaction time.    
However, duty-cycled IoT systems do not typically require separate modes for mandatory services and extra services, as an optimized system only runs the software required to complete its task.
When device features, such as applying system updates, are not utilized during every duty-cycle, they are easily loaded after the boot up is completed. 
Furthermore, modern Linux kernels support dynamic kernel module loading using the \texttt{modprobe} utility.

\subsubsection{Suspend-related Improvements}
Another approach to improve boot up time is the usage of a suspend or hibernation mechanism.
In \cite{park2006linux}, the standard \textit{hibernate-resume} approach is used to optimize the boot up time of a digital camera.
During the \textit{system suspend} phase, current system information such as processor registers, I/O map information, and runtime variables (both global and local), are stored in RAM before the device powers off.
When the system receives certain power events, the \textit{resume} operation is started.
The previous state of the device is restored from RAM, resulting in a faster initialization time than a regular boot up.
It is important to note that the RPi platform's hardware does not support hibernation or suspension, as the SoCs do not support modern power management features \cite{BroadcomSpecSheet}.

In \cite{kaminaga2006improving} and \cite{joe2011bootup}, methods for \textit{snapshot boot techniques} are discussed with respect to boot up time improvement on a device running a Linux-based OS.
These methods utilize a snapshot image created at boot up time and stored on a disk or in reserved flash memory.
For subsequent boot ups, the device loads the suspended image instead of stepping through the standard kernel initialization process.
As opposed to a standard \textit{hibernate-resume} operation, this snapshot image is generated only once and then reused.
A major disadvantage of snapshot images is the considerable amount time required for image generation, verification, and storage.
Also, storing a full image in addition to the original system scripts and binaries necessary for image generation requires a considerable amount of disk space, which may not be available on many IoT platforms.
If the image is corrupted (due to power loss, for examples), the device may become unrecoverable if a fallback image is unavailable.
To mitigate these issues, the authors in \cite{baik2010boosting} proposed a dual-image system, where some essential services and initialization are kept in the first image, and less critical functions are loaded by the second image, referred to as the \textit{essential-snapshot-image} and \textit{supplementary-snapshot-image}, respectively.



\subsubsection{Bootloader Optimization}
The bootloader is responsible for initializing the hardware and loading the kernel.  
In most cases, the kernel is compressed to save space and the bootloader must decompress it before use.
The authors in  \cite{chung2007study} analyzed the time required by the bootloader and compared the performance of several root filesystems for fast boot up time.
Decompressing the kernel with \texttt{gzip-cheksum} on their system resulted in a bootloader time of 16s.
Conversely, using \texttt{gzip-nochecksum} required only 12s, with a reduction of 3.8s.
While the actual decompression of the image took 2.79s, verifying the checksum required an extra second on their device.
In addition, the cost of calculating the checksum for the RAM disk was 2.78s.
They found that storing the image in a decompressed format can circumvent this process.
These results are noteworthy in the realm of embedded Linux, where processors may be slower than the RPi's processor and the onboard flash is more responsive than the SDC used by the RPi.
However, most general-purpose Linux systems can decompress the kernel faster than the storage medium can read the uncompressed alternative.
Therefore, it is important to balance the speed of the processor and the I/O read speed before seeking performance gains from an uncompressed kernel.

Similar to \textit{Application XIP}, \textit{Kernel XIP} (Kernel Execute-In-Place) \cite{Bird2010MethodsTI} is a popular technique used for optimizing bootloader initialization.
In a typical boot sequence, the kernel is decompressed either during or just after it is loaded into memory.
XIP enables direct execution of kernel instructions directly from ROM or flash memory. 
However, this method requires the kernel be stored in an uncompressed format, thereby requiring more storage space.
The RPi platform does not support \texttt{Kernel XIP} due to hardware constraints.

\subsubsection{Kernel Space Optimization}

In \cite{chung2007study}, the authors show the effect of removing unnecessary kernel modules and bundling multiple modules into a single module on decreasing kernel loading time.  
In another work \cite{Bird2010MethodsTI}, the authors evaluate the effect of disabling kernel print statements to prevent bottlenecks caused by streaming messages to the console.
The \texttt{quiet} option in the kernel configuration changes the logging level of print statements to 4, which suppresses the output of regular (non-emergency) messages.

\subsection{Energy-Efficient Software Development}
\label{energy_eff_sw_dev}
The importance of energy efficient software development has been attracting a lot of attention from both industry and academia \cite{procaccianti2016empirical,pinto2016comprehensive,georgiou2018your}.
The importance, in particular, is revealed by \cite{pinto2014mining}, which studies the popularity of energy-efficient software design by developers.
Specifically, for mobile application development, the literature proposes various approaches and tools to study and improve energy consumption.
For example, \cite{li2013calculating} combines empirical power measurement with statistical modeling to offer energy consumption information at the source line level.
The authors in \cite{hao2012estimating} propose a technique that first determines the execution trace of the code and then applies a set of cost functions to estimate the energy consumption of the execution path.
The authors in \cite{hindle2015green} study the impact of software maintenance and update on energy consumption.
Although these solutions do not directly address the energy consumption and overhead of duty-cycled systems from the boot up and shutdown point of views, they are complementary to our work and can be used to enhance the performance of these systems further. 
For example, developers can first use these techniques to improve the energy utilization of an user application and identify the energy bugs. 
Then Pallex can be employed to parallelize the execution of the user application's stages with the userspace initialization phase.


\section{Conclusion}
\label{section::conclusion} 

The increasing number of Linux-based IoT devices used for edge and fog computing necessitates the adoption of duty-cycling mechanisms to reduce the energy consumption of these devices.
To this end, profiling and improving the operation of userspace initialization offers techniques that can be easily adopted by users.
In this work, we presented a thorough study of the Linux boot up process, in particular the effects of unit activation on the duration and energy consumption of the boot up and shutdown phases.
We showed that although some units cannot be disabled without compromising system stability, duty-cycling performance is significantly enhanced by application-specific unit configuration.
Our studies also showed that there is no effect on boot up performance when up to 95\% of the SDC capacity is utilized. 
However, using a faster SDC results in a slightly shorter, more energy efficient duty-cycle.
After analyzing the resource utilization of the RPi3 and RPiZW during the boot up process, we showed that user applications can be executed in parallel with the userspace initialization phase to reduce the energy consumption of each duty-cycle.
To this end, we proposed Pallex, a parallel execution framework which relies on \texttt{systemd} and Unix Domain Sockets to break a user application into multiple phases.
Our evaluations show up to a 31\% reduction in energy consumption and up to a 30\% enhancement in lifetime when Pallex is applied to various IoT application scenarios.
Our studies also reveal the trade-off between processing power and current consumption.
Although the current draw of the RPiZW is lower than that of the RPi3, this paper confirms the RPi3 platform is more suitable for duty-cycled applications.
This is because the RPi3 parallelizes userspace initialization and user application processes across multiple cores, resulting in lower energy consumption by shortening processing duration.

Some potential areas of future work are as follows:
Although the studies of this paper revealed the significant effects of quad-core and single-core SoCs on duty-cycling performance, extending these observations and profiling the performance of other COTS Linux-based boards is of interest.
From the shutdown point of view, developing a model to predict the probability of SDC corruption based on factors such as capacity, I/O rate, and duty-cycling frequency enables users to choose the best shutdown mechanism available without compromising system reliability.
Regarding Pallex, although we have provided guidelines to simplify its applicability to other IoT scenarios, it would be helpful to develop a program to analyze user application code and break it into stages based on the tasks it performs and the dependencies of those tasks.
Finally, unit configuration and Pallex can be integrated with bootloader and kernel-level optimization mechanisms to further enhance performance.


\section*{Acknowledgment}
This research has been partially supported by the Santa Clara Valley Water District research grant SCWD02.
This project involves the development of a flood monitoring system where Linux-based wireless systems, which rely on solar or battery power, capture images for analysis using machine learning to classify and report the debris carried by rivers and streams.

\appendices

\section{List of Units in Raspbian Stretch Lite (RSL)}
\label{appendix:services}
In this section, we present an overview of the units available in RSL.
All units except those in the EU category can be disabled if they are not necessary for the application scenario being considered.

\begin{itemize}

\subsection{Essential Units (EU)}
\label{app_es}


\item [--] \textbf{\texttt{boot.mount}}: 
	This unit helps \texttt{systemd} resolve dependency trees for units that depend on mounting \texttt{/boot} before activation.

\item [--] \textbf{\texttt{dev-mmcblk0p2.device}}: 
This unit brings the root partition on the SDC into the scope of \texttt{systemd} so that units that require the root partition's mount to finish before activation can resolve their dependencies properly.

\item [--] \textbf{\texttt{dev-mqueue.mount}}: 
This unit informs \texttt{systemd} when the POSIX message queues for internal system messages is ready.

\item [--] \textbf{\texttt{kmod.service}}: 
This service contains \texttt{modprobe}, which is used for loading and unloading kernel modules. 

\item [--] \textbf{\texttt{kmod-static-nodes.service}}:
This service creates a list of required static modules for the loaded kernel. 


\item [--] \textbf{\texttt{run-rpc\_pipefs.mount}}:
This unit directs \texttt{systemd} on how to mount the RAM-based \texttt{pipefs}, which is used every time a process is forked or a pipe (``|'') is used.
	
\item [--] \textbf{\texttt{sys-kernel-debug.mount}}: 
Similar to \texttt{dev-mmcblk0p2.device}, this unit helps \texttt{systemd} resolve dependencies correctly. 
The actual mounting of \texttt{debugfs} occurs within \texttt{udev}, which is the daemon that detects hardware changes. 

\item [--] \textbf{\texttt{sys-kernel-config.mount}}: 
This unit prevents the system from reaching \texttt{sysinit.target} until the kernel configuration parameters are fully loaded into the kernel from the Configuration File System (\texttt{configfs}). 

\item [--] \textbf{\texttt{systemd-fsck\@.service}} and \textbf{\texttt{systemd-fsck-root.service}}:
These services run \texttt{fsck} on each partition to ensure file system consistency.
This is an important step, and does not run every time unless there are problems detected on the SDC.


\item [--] \textbf{\texttt{systemd-journald.service}}:
Many programs rely on \texttt{journald} for logging output, including the kernel (through \texttt{kmsg}).
Therefore, it should not be disabled. 
However, in order to speed up its initialization, it may be useful to lower the size limit of the journal logs since a dependency, \texttt{systemd-journal-flush.service}, must rotate this log file on initialization.

\item [--] \textbf{\texttt{systemd-modules-load.service}}: 
This service starts early in the userspace initialization phase to load static kernel modules. 

\item [--] \textbf{\texttt{systemd-remount-fs.service}}: 
In the beginning of the userspace initialization phase, this service mounts the necessary API filesystems for the kernel (such as \texttt{/proc}, \texttt{/sys}, or \texttt{/dev}) to a RAM disk.

\item [--] \textbf{\texttt{systemd-random-seed.service}}: 
This service loads the random seed and saves it at shutdown to enable the device to generate a new value when the system restarts.

\item [--] \textbf{\texttt{systemd-sysctl.service}}: 
By loading kernel configurations, this service enables \texttt{systemctl} to perform as expected.

\item [--] \textbf{\texttt{systemd-udevd.service}}: 
This service initializes \texttt{udev}, a daemon that listens to kernel \texttt{uevents} and matches them against specified rules, to run scripts.
For example, it can load drivers when a new device is attached, or mount a USB drive when it is plugged in.

\item [--] \textbf{\texttt{systemd-udev-trigger.service}}: 
Devices plugged in before the system is powered on might not generate the kernel messages necessary for \texttt{udev} to discover them.
This service probes and detects devices that \texttt{udev} would not normally discover.


\item [--] \textbf{\texttt{sudo.service}}: 
This service clears cached sudo privilege escalations to enforce user re-authentication after every reboot.

\item [--] \textbf{\texttt{systemd-tmpfiles-setup.service}} and \textbf{\texttt{systemd-tmpfiles-setup-dev.service}}: 
Mount \texttt{/tmp} and delete the old files.
These services also create any files that are specified by user-provided configuration. 

\item [--] \textbf{\texttt{systemd-rfkill.service}}: 
This service restores the \texttt{rfkill} state at the beginning of userspace initialization to ensure it matches the status saved before shutdown.
Therefore, if the wireless peripherals (typically WiFi or Bluetooth) had \texttt{rfkill} preventing their use before shutdown, they will remain disabled on reboot.

\item [--] \textbf{\texttt{systemd-update-utmp.service}} and \textbf{\texttt{systemd-update-utmp-runlevel.service}}: Record and manage the system uptime, the logged-in users, and users' log-in method (such as ssh, tty, and serial port).

\item [--] \textbf{\texttt{systemd-user-sessions.service}}: 
This service enables user log-in and denies log-in attempts after the shutdown signal has been sent. 
If a device in the field does not require log-in capabilities, this service does not need to be started automatically. 
For example, it can be started by a helper program when a GPIO pin is pulled high.

\item [--] \textbf{\texttt{systemd-logind.service}}: 
This service is responsible for tasks such as user session management, processor usage quotas, and device access management.

\subsection{Networking-related Services (NRS)}
\label{appendix_nrs}

\item [--] \textbf{\texttt{avahi-daemon.service}}: 
This service enables programs to discover and publish services and hosts running on a local network.
Note that this service can significantly slow the speed of the \texttt{ifdown} command and therefore the shutdown process if not completely uninstalled.
Unless necessary for the user application, \texttt{avahi-daemon} should be uninstalled.

\item [--] \textbf{\texttt{bluetoothd.service}}: 
Daemon for controlling the Bluetooth interface. 
\texttt{Bluez}, \texttt{bluetoothctl}, and many other Bluetooth-related utilities communicate through this daemon.

\item [--] \textbf{\texttt{dhcpcd.service}}: 
The daemon responsible for managing the DHCP protocol on all targeted network interfaces.
This service can be disabled if the device does not require a network connection or is guaranteed a static IP address.
The duration of IP allocation depends on external factors including link quality and the load of access point.

\item [--] \textbf{\texttt{hciuart.service}}: 
This is responsible for initializing the HCI bluetooth interface.
HCI stands for "Host-to-Controller-Interface" and it is controlled over a serial UART interface.

\item [--] \textbf{\texttt{networking.service}}: 
Completes the configuration of WiFi and Ethernet interfaces based on the settings available in the \texttt{/etc/network/interfaces} configuration file.

\item [--] \textbf{\texttt{nfs-config.service}}: 
This service, along with \texttt{nfs-common.service}, loads configuration details applicable to Network File Systems (NFS).

\item [--] \textbf{\texttt{rsyncd.service}}: 
Daemon that listens on port 873 for incoming \texttt{rsync} file transfer requests. 
\texttt{rsync} is used for efficiently transferring and synchronizing files across computer systems.


\item [--]\textbf{\texttt{rpcbind.service}}:
This service accepts requests for Remote Procedure Calls (RPC) and binds them to TCP ports for access and control.

\item [--] \textbf{\texttt{sshd.service}}: 
This service belongs to the \texttt{OpenSSH} package.
It runs in the background to listen for and accept or deny incoming ssh connections according to a user-defined configuration file.



\item [--] \textbf{\texttt{systemd-hostnamed.service}}: 
This service can be used to control the hostname and related metadata by user programs.  

\item [--] \textbf{\texttt{systemd-networkd.service}}: 	
Brings up the system's network manager and provides it with discovered networks, both physical and virtual.

\item [--] \textbf{\texttt{systemd-resolved.service}}: 
	Provides local DNS resolution for namespaces such as \texttt{localhost} or those added by the user to overlay DNS provided by an external source.

\item [--] \textbf{\texttt{systemd-timesyncd.service}}: 
	Used for time synchronization across the network.




\subsection{Memory Management Services} 
\item [--] \textbf{\texttt{dphys-swapfile.service}}: 
This service initializes, mounts, unmounts, and deletes swap files on the SDC. 
If the available RAM is enough for the user application, then disabling this service results in a performance enhancement in terms of faster boot up time and prolonged SDC lifetime.
This service is usually required when the user application involves loading large machine learning models and data sets.







\subsection{I/O-related Services}
\item [--] \textbf {\texttt{alsa-utils.service}}: Represents the tools relating to the Advanced Linux Sound Architecture (ALSA).

\item [--] \textbf{\texttt{alsa-restore.service}}: 
Initializes and restores the last state of the RPi's onboard soundcard.



\subsection{Miscellaneous Units}

\item [--] \textbf{\texttt{fake-hwclock.service}}: 
This service saves the current time to a file at shutdown and loads it at boot up time.
Without this service, the RPi is unaware of the current epoch time until it establishes a network connection. 
An incorrect epoch value may cause some files to appear as if they are edited in the future.

\item [--] \textbf{\texttt{plymouth.service}}: 
Provides a flicker-free graphical boot up process.
Other related services include \texttt{plymouth-quit.service}, \texttt{plymouth-quit-wait.service}, \texttt{plymouth-start. service}, and \texttt{plymouth-read-write.service}.

\item [--] \textbf{\texttt{raspi-config.service}}: 	
This service loads configuration changes made by the user such as processor governance, display overscan, and filesystem partition expansions, and applies them on reboot.

\item [--] \textbf{\texttt{rsyslog.service}}: 
Tools for log processing and conversion.

\item [--] \textbf{\texttt{console-setup.service}}: 
Configures the fonts, screen resolution, keyboard layout, etc., for virtual tty terminals.

\end{itemize}

\ifCLASSOPTIONcaptionsoff
  \newpage
\fi



%

\bibliographystyle{IEEEtran}
\bibliography{references}

\begin{thebibliography}{10}
\providecommand{\url}[1]{#1}
\csname url@samestyle\endcsname
\providecommand{\newblock}{\relax}
\providecommand{\bibinfo}[2]{#2}
\providecommand{\BIBentrySTDinterwordspacing}{\spaceskip=0pt\relax}
\providecommand{\BIBentryALTinterwordstretchfactor}{4}
\providecommand{\BIBentryALTinterwordspacing}{\spaceskip=\fontdimen2\font plus
\BIBentryALTinterwordstretchfactor\fontdimen3\font minus
  \fontdimen4\font\relax}
\providecommand{\BIBforeignlanguage}[2]{{%
\expandafter\ifx\csname l@#1\endcsname\relax
\typeout{** WARNING: IEEEtran.bst: No hyphenation pattern has been}%
\typeout{** loaded for the language `#1'. Using the pattern for}%
\typeout{** the default language instead.}%
\else
\language=\csname l@#1\endcsname
\fi
#2}}
\providecommand{\BIBdecl}{\relax}
\BIBdecl

\bibitem{delforge2016slashing}
P.~Delforge, ``Slashing energy use in computers and monitors while protecting
  our wallets, health, and planet.''\hskip 1em plus 0.5em minus 0.4em\relax
  Natural Resources Defense Council, 2016.

\bibitem{chu2012opportunities}
S.~Chu and A.~Majumdar, ``Opportunities and challenges for a sustainable energy
  future,'' \emph{nature}, vol. 488, no. 7411, p. 294, 2012.

\bibitem{wang2015energy}
Z.~Wang, Y.~Liu, Y.~Sun, Y.~Li, D.~Zhang, and H.~Yang, ``\relax{An
  energy-efficient heterogeneous dual-core processor for Internet of Things},''
  in \emph{IEEE International Symposium on Circuits and Systems (ISCAS)}.\hskip
  1em plus 0.5em minus 0.4em\relax IEEE, 2015, pp. 2301--2304.

\bibitem{ueki2015low}
M.~Ueki, K.~Takeuchi, T.~Yamamoto, A.~Tanabe, N.~Ikarashi, M.~Saitoh,
  T.~Nagumo, H.~Sunamura, M.~Narihiro, K.~Uejima \emph{et~al.}, ``Low-power
  embedded reram technology for iot applications,'' in \emph{Symposium on VLSI
  Technology}.\hskip 1em plus 0.5em minus 0.4em\relax IEEE, 2015, pp.
  T108--T109.

\bibitem{CYW43907}
\BIBentryALTinterwordspacing
{\relax Cypress Semiconductor}. {\relax CYW43907: IEEE 802.11 a/b/g/n SoC with
  an Embedded Applications Processor}. [Online]. Available:
  \url{http://www.cypress.com/file/298236/download}
\BIBentrySTDinterwordspacing

\bibitem{baccelli2013riot}
E.~Baccelli, O.~Hahm, M.~Gunes, M.~Wahlisch, and T.~C. Schmidt, ``\relax{RIOT
  OS: Towards an OS for the Internet of Things},'' in \emph{IEEE Conference on
  Computer Communications Workshops (INFOCOM Workshops)}.\hskip 1em plus 0.5em
  minus 0.4em\relax IEEE, 2013, pp. 79--80.

\bibitem{levis2005tinyos}
P.~Levis, S.~Madden, J.~Polastre, R.~Szewczyk, K.~Whitehouse, A.~Woo, D.~Gay,
  J.~Hill, M.~Welsh, E.~Brewer \emph{et~al.}, ``\relax{TinyOS: An operating
  system for sensor networks},'' in \emph{Ambient intelligence}.\hskip 1em plus
  0.5em minus 0.4em\relax Springer, 2005, pp. 115--148.

\bibitem{ThreadX}
\BIBentryALTinterwordspacing
(2018) \relax{ThreadX RTOS Real-Time Operating System}. [Online]. Available:
  \url{https://rtos.com/solutions/threadx/real-time-operating-system/}
\BIBentrySTDinterwordspacing

\bibitem{FreeRTOS}
\BIBentryALTinterwordspacing
(2018) \relax{The FreeRTOS Kernel}. [Online]. Available:
  \url{https://www.freertos.org}
\BIBentrySTDinterwordspacing

\bibitem{jones2001survey}
C.~E. Jones, K.~M. Sivalingam, P.~Agrawal, and J.~C. Chen, ``A survey of energy
  efficient network protocols for wireless networks,'' \emph{wireless
  networks}, vol.~7, no.~4, pp. 343--358, 2001.

\bibitem{dezfouli2017rewimo}
B.~Dezfouli, M.~Radi, and O.~Chipara, ``Rewimo: a real-time and reliable
  low-power wireless mobile network,'' \emph{ACM Transactions on Sensor
  Networks (TOSN)}, vol.~13, no.~3, p.~17, 2017.

\bibitem{dezfouli2015dicsa}
B.~Dezfouli, M.~Radi, K.~Whitehouse, S.~A. Razak, and T.~Hwee-Pink,
  ``\relax{DICSA: Distributed and concurrent link scheduling algorithm for data
  gathering in wireless sensor networks},'' \emph{Ad Hoc Networks}, vol.~25,
  pp. 54--71, 2015.

\bibitem{zoican2012lwip}
S.~Zoican and M.~Vochin, ``\relax{LwIP stack protocol for embedded sensors
  network},'' in \emph{9th International Conference on Communications}.\hskip
  1em plus 0.5em minus 0.4em\relax IEEE, 2012, pp. 221--224.

\bibitem{pinto2014mining}
G.~Pinto, F.~Castor, and Y.~D. Liu, ``Mining questions about software energy
  consumption,'' in \emph{Proceedings of the 11th Working Conference on Mining
  Software Repositories}.\hskip 1em plus 0.5em minus 0.4em\relax ACM, 2014, pp.
  22--31.

\bibitem{ietf_edge}
\BIBentryALTinterwordspacing
J.~Hong, Y.-G. Hong, and J.-S. Youn, ``\relax{Problem Statement of IoT
  integrated with Edge Computing},'' 2018. [Online]. Available:
  \url{https://tools.ietf.org/html/draft-hong-iot-edge-computing-00}
\BIBentrySTDinterwordspacing

\bibitem{Griffiths_2018}
E.~Griffiths, S.~Assana, and K.~Whitehouse, ``Privacy-preserving image
  processing with binocular thermal cameras,'' \emph{Proc. ACM Interact. Mob.
  Wearable Ubiquitous Technologies}, vol.~1, no.~4, pp. 133:1--133:25, 2018.

\bibitem{kelly2016internet}
R.~Kelly, ``Internet of things data to top 1.6 zettabytes by 2022,''
  \emph{Campus Technology}, vol.~9, pp. 1536--1233, 2016.

\bibitem{chiang2016fog}
M.~Chiang and T.~Zhang, ``\relax{Fog and IoT: An overview of research
  opportunities},'' \emph{IEEE Internet of Things Journal}, vol.~3, no.~6, pp.
  854--864, 2016.

\bibitem{iot_survey}
\BIBentryALTinterwordspacing
\relax {Eclipse Foundation}, ``\relax{Key Trends from the IoT Developer Survey
  2018}.'' [Online]. Available:
  \url{https://blogs.eclipse.org/post/benjamin-cabe/key-trends-iot-developer-survey-2018}
\BIBentrySTDinterwordspacing

\bibitem{kaup2014powerpi}
F.~Kaup, P.~Gottschling, and D.~Hausheer, ``\relax{PowerPi: Measuring and
  modeling the power consumption of the Raspberry Pi},'' in \emph{39th
  Conference on Local Computer Networks (LCN)}.\hskip 1em plus 0.5em minus
  0.4em\relax IEEE, 2014, pp. 236--243.

\bibitem{morabito2017virtualization}
R.~Morabito, ``Virtualization on internet of things edge devices with container
  technologies: a performance evaluation,'' \emph{IEEE Access}, vol.~5, pp.
  8835--8850, 2017.

\bibitem{vujovic2014raspberry}
V.~Vujovic and M.~Maksimovic, ``\relax{Raspberry Pi as a Wireless Sensor node:
  Performances and constraints},'' in \emph{37th International Convention on
  Information and Communication Technology, Electronics and Microelectronics
  (MIPRO)}.\hskip 1em plus 0.5em minus 0.4em\relax IEEE, 2014, pp. 1013--1018.

\bibitem{fisher2015open}
R.~Fisher, L.~Ledwaba, G.~Hancke, and C.~Kruger, ``Open hardware: A role to
  play in wireless sensor networks?'' \emph{Sensors}, vol.~15, no.~3, pp.
  6818--6844, 2015.

\bibitem{dezfouli2015modeling}
B.~Dezfouli, M.~Radi, S.~A. Razak, T.~Hwee-Pink, and K.~A. Bakar, ``Modeling
  low-power wireless communications,'' \emph{Journal of Network and Computer
  Applications}, vol.~51, pp. 102--126, 2015.

\bibitem{chung2007study}
K.~H. Chung, M.~S. Choi, and K.~S. Ahn, ``\relax{A study on the packaging for
  fast boot-up time in the embedded Linux},'' in \emph{13th International
  Conference on Embedded and Real-Time Computing Systems and Applications
  (RTCSA)}.\hskip 1em plus 0.5em minus 0.4em\relax IEEE, 2007, pp. 89--94.

\bibitem{googleDebian}
\BIBentryALTinterwordspacing
C.~Villegas. (2006) \relax{Improve the Debian boot process}. [Online].
  Available: \url{http://bootdebian.blogspot.com}
\BIBentrySTDinterwordspacing

\bibitem{kaminaga2006improving}
H.~Kaminaga, ``Improving linux startup time using software resume (and other
  techniques),'' in \emph{Linux Symposium}, 2006, p.~17.

\bibitem{Villegas2006ImproveTD}
\BIBentryALTinterwordspacing
C.~Villegas and P.~Reinholdtsen, ``State-of the-art in the boot
  process.''\hskip 1em plus 0.5em minus 0.4em\relax Google Summer of Code,
  2006. [Online]. Available:
  \url{https://pdfs.semanticscholar.org/a171/696ddb41ba8aad53cdcbb6aba1c4547aa80e.pdf}
\BIBentrySTDinterwordspacing

\bibitem{Bird2010MethodsTI}
T.~R. Bird, ``\relax{Methods to Improve Bootup Time in Linux},'' in
  \emph{Proceedings of the Linux Symposium}, 2004.

\bibitem{singh2011optimizing}
G.~Singh, K.~Bipin, and R.~Dhawan, ``Optimizing the boot time of android on
  embedded system,'' in \emph{15th International Symposium on Consumer
  Electronics (ISCE)}.\hskip 1em plus 0.5em minus 0.4em\relax IEEE, 2011, pp.
  503--508.

\bibitem{jo2009optimizing}
H.~Jo, H.~Kim, J.~Jeong, J.~Lee, and S.~Maeng, ``Optimizing the startup time of
  embedded systems: a case study of digital tv,'' \emph{IEEE Transactions on
  Consumer Electronics}, vol.~55, no.~4, 2009.

\bibitem{upton2016learning}
E.~Upton, J.~Duntemann, R.~Roberts, B.~Everard, and T.~Mamtora,
  \emph{\relax{Learning Computer Architecture with Raspberry Pi}}.\hskip 1em
  plus 0.5em minus 0.4em\relax John Wiley \& Sons, 2016.

\bibitem{BroadcomSpecSheet}
\BIBentryALTinterwordspacing
``\relax{BCM2835 ARM Peripherals}.'' [Online]. Available:
  \url{https://www.raspberrypi.org/app/uploads/2012/02/BCM2835-ARM-Peripherals.pdf}
\BIBentrySTDinterwordspacing

\bibitem{systemdInit}
\BIBentryALTinterwordspacing
(2018) \relax{systemd System and Service Manager}. [Online]. Available:
  \url{https://www.freedesktop.org/wiki/Software/systemd/}
\BIBentrySTDinterwordspacing

\bibitem{Gorauskas_2015}
J.~Gorauskas, ``\relax{Managing Services in Linux: Past, Present and Future},''
  \emph{Linux J.}, vol. 2015, no. 251, 2015.

\bibitem{systemd_unit}
\BIBentryALTinterwordspacing
``\relax{systemd.unit},'' 2018. [Online]. Available:
  \url{https://www.freedesktop.org/software/systemd/man/systemd.unit.html}
\BIBentrySTDinterwordspacing

\bibitem{LinuxKernelDev}
R.~Love, \emph{Linux Kernel Development}, 3rd~ed.\hskip 1em plus 0.5em minus
  0.4em\relax Addison-Wesley, 2010.

\bibitem{dezfouli2018empiot}
B.~Dezfouli, I.~Amirtharaj, and C.-C. Li, ``\relax{EMPIOT: An energy
  measurement platform for wireless IoT devices},'' \emph{Journal of Network
  and Computer Applications}, vol. 121, pp. 135 -- 148, 2018.

\bibitem{bcmdriver}
\BIBentryALTinterwordspacing
``\relax{C library for Broadcom BCM 2835}.'' [Online]. Available:
  \url{http://www.airspayce.com/mikem/bcm2835/}
\BIBentrySTDinterwordspacing

\bibitem{wiringpi}
\BIBentryALTinterwordspacing
``\relax{Wiring Pi: GPIO Interface library for the Raspberry Pi}.'' [Online].
  Available: \url{http://wiringpi.com}
\BIBentrySTDinterwordspacing

\bibitem{kth_dmm_7510}
\BIBentryALTinterwordspacing
{\relax Tektronix}. \relax {DMM7510 7$\frac{1}{2}$ Digit Graphical Sampling
  Multimeter}. [Online]. Available:
  \url{https://www.tek.com/tektronix-and-keithley-digital-multimeter/dmm7510}
\BIBentrySTDinterwordspacing

\bibitem{Jiang2007}
X.~Jiang, P.~Dutta, D.~Culler, and I.~Stoica, ``{Micro Power Meter for Energy
  Monitoring of Wireless Sensor Networks at Scale},'' 2007, p. 186.

\bibitem{Stathopoulos2008}
T.~Stathopoulos, D.~McIntire, and W.~J. Kaiser, ``{The energy endoscope:
  Real-time detailed energy accounting for wireless sensor nodes},''
  \emph{Proceedings of International Conference on Information Processing in
  Sensor Networks (IPSN'08)}, pp. 383--394, 2008.

\bibitem{Duttac2008}
P.~Dutta, M.~Feldmeier, J.~Paradiso, and D.~Culler, ``Energy metering for free:
  Augmenting switching regulators for real-time monitoring,'' in
  \emph{Proceedings of the 7th International Conference on Information
  Processing in Sensor Networks (IPSN'08)}, 2008, pp. 283--294.

\bibitem{Andersen2009}
J.~Andersen and M.~T. Hansen, ``{Energy Bucket: A Tool for Power Profiling and
  Debugging of Sensor Nodes},'' in \emph{Proceedings of Third International
  Conference on Sensor Technologies and Applications (SENSORCOMM'09)}.\hskip
  1em plus 0.5em minus 0.4em\relax IEEE, 2009, pp. 132--138.

\bibitem{Zhou2013a}
R.~Zhou and G.~Xing, ``{Nemo: A high-fidelity noninvasive power meter system
  for wireless sensor networks},'' \emph{Proceedings of the ACM/IEEE
  International Conference on Information Processing in Sensor Networks
  (IPSN'13)}, pp. 141--152, 2013.

\bibitem{Naderiparizi2016}
S.~Naderiparizi, A.~N. Parks, F.~S. Parizi, and J.~R. Smith, ``{$\mu$Monitor:
  In-situ energy monitoring with microwatt power consumption},'' in
  \emph{Proceedings of the IEEE International Conference on RFID
  (RFID'16)}.\hskip 1em plus 0.5em minus 0.4em\relax IEEE, may 2016, pp. 1--8.

\bibitem{Haratcherev2008}
I.~Haratcherev, G.~Halkes, and T.~Parker, ``{PowerBench: A Scalable Testbed
  Infrastructure for Benchmarking Power Consumption},'' in \emph{Proceedings of
  the International Workshop on Sensor Network Engineering (IWSNE'08)}, 2008,
  pp. 37--44.

\bibitem{Trathnigg2008}
T.~Trathnigg, M.~J{\"u}rgen, and R.~Weiss, ``A low-cost energy measurement
  setup and improving the accuracy of energy simulators for wireless sensor
  networks,'' in \emph{Proceedings of the workshop on Real-world wireless
  sensor networks}, 2008, pp. 31--35.

\bibitem{Zhu2013b}
N.~Zhu and I.~O'Connor, ``Energy measurements and evaluations on high data rate
  and ultra low power wsn node,'' in \emph{10th IEEE International Conference
  on Networking, Sensing and Control (ICNSC)}, 2013, pp. 232--236.

\bibitem{Hartung2016}
R.~Hartung, U.~Kulau, and L.~Wolf, ``{Distributed energy measurement in WSNs
  for outdoor applications},'' 2016, pp. 1--9.

\bibitem{Potsch2017}
A.~P{\"o}tsch, A.~Berger, and A.~Springer, ``Efficient analysis of power
  consumption behaviour of embedded wireless iot systems,'' in
  \emph{Proceedings of the Instrumentation and Measurement Technology
  Conference (I2MTC)}, 2017, pp. 1--6.

\bibitem{Gomez2012a}
K.~Gomez, R.~Riggio, T.~Rasheed, D.~Miorandi, and F.~Granelli, ``{Energino: a
  Hardware and Software Solution for Energy Consumption Monitoring},'' in
  \emph{Proceedings of the International Workshop on Wireless Network
  Measurements (WiOpt'12)}, 2012, pp. 311 -- 317.

\bibitem{Keranidis2014b}
S.~Keranidis, G.~Kazdaridis, V.~Passas, T.~Korakis, I.~Koutsopoulos, and
  L.~Tassiulas, ``{NITOS Energy Monitoring Framework: Real Time Power
  Monitoring in Experimental Wireless Network Deployments},'' \emph{SIGMOBILE
  Mob. Comput. Commun. Rev.}, vol.~18, no.~1, pp. 64--74, 2014.

\bibitem{systemctlService}
\BIBentryALTinterwordspacing
``systemctl,'' 2017. [Online]. Available:
  \url{https://www.freedesktop.org/software/systemd/man/systemctl.html}
\BIBentrySTDinterwordspacing

\bibitem{gawk}
\BIBentryALTinterwordspacing
``\relax{The GNU Awk User's Guide}.'' [Online]. Available:
  \url{https://www.gnu.org/software/gawk/manual/gawk.html}
\BIBentrySTDinterwordspacing

\bibitem{iostat}
\BIBentryALTinterwordspacing
S.~Godard, ``iostat,'' 2018. [Online]. Available:
  \url{http://man7.org/linux/man-pages/man1/iostat.1.html}
\BIBentrySTDinterwordspacing

\bibitem{RPiCam}
\BIBentryALTinterwordspacing
(2018) \relax{Camera Module (v2)}. [Online]. Available:
  \url{https://www.raspberrypi.org/documentation/hardware/camera/}
\BIBentrySTDinterwordspacing

\bibitem{pi_camera_cmos}
\BIBentryALTinterwordspacing
(2018) \relax{IMX219PQ: Diagonal 4.6mm 8.08M-Effective Pixel Color CMOS Image
  Sensor}. [Online]. Available:
  \url{https://www.sony-semicon.co.jp/products_en/new_pro/april_2014/imx219_e.html}
\BIBentrySTDinterwordspacing

\bibitem{DDTool}
\BIBentryALTinterwordspacing
S.~K. Paul~Rubin, David~MacKenzie, ``\relax{dd: convert and copy a file},''
  2018. [Online]. Available:
  \url{http://man7.org/linux/man-pages/man1/dd.1.html}
\BIBentrySTDinterwordspacing

\bibitem{gupta2009dftl}
A.~Gupta, Y.~Kim, and B.~Urgaonkar, ``\relax{DFTL: A Flash Translation Layer
  Employing Demand-based Selective Caching of Page-level Address Mappings},''
  in \emph{Proceedings of the 14th International Conference on Architectural
  Support for Programming Languages and Operating Systems}.\hskip 1em plus
  0.5em minus 0.4em\relax ACM, 2009, pp. 229--240.

\bibitem{unixDomainSocket}
\BIBentryALTinterwordspacing
``\relax{Linux Programmer's Manual, Socket},'' 2018. [Online]. Available:
  \url{http://man7.org/linux/man-pages/man2/socket.2.html}
\BIBentrySTDinterwordspacing

\bibitem{bovet2005understanding}
D.~P. Bovet and M.~Cesati, \emph{Understanding the Linux Kernel: from I/O ports
  to process management}.\hskip 1em plus 0.5em minus 0.4em\relax " O'Reilly
  Media, Inc.", 2005.

\bibitem{picamera}
\BIBentryALTinterwordspacing
D.~Jones. \relax{Python interface to the Raspberry Pi camera module}. [Online].
  Available: \url{https://picamera.readthedocs.io/en/release-1.13}
\BIBentrySTDinterwordspacing

\bibitem{altman1992introduction}
N.~S. Altman, ``An introduction to kernel and nearest-neighbor nonparametric
  regression,'' \emph{The American Statistician}, vol.~46, no.~3, pp. 175--185,
  1992.

\bibitem{scikit-learn}
\BIBentryALTinterwordspacing
``scikit-learn: machine learning in python.'' [Online]. Available:
  \url{https://scikit-learn.org/}
\BIBentrySTDinterwordspacing

\bibitem{georgiou2018your}
S.~Georgiou, M.~Kechagia, P.~Louridas, and D.~Spinellis, ``What are your
  programming language's energy-delay implications?'' in \emph{Proceedings of
  the 15th International Conference on Mining Software Repositories
  (MSR)}.\hskip 1em plus 0.5em minus 0.4em\relax ACM, 2018, pp. 303--313.

\bibitem{brodowski2015cpu}
\BIBentryALTinterwordspacing
D.~Brodowski, N.~Golde, R.~J.~Wysocki, and V.~Kumar, ``\relax{CPU frequency and
  voltage scaling code in the Linux (TM) kernel}.'' [Online]. Available:
  \url{https://www.kernel.org/doc/Documentation/cpu-freq/governors.txt}
\BIBentrySTDinterwordspacing

\bibitem{hindle2015green}
A.~Hindle, ``Green mining: a methodology of relating software change and
  configuration to power consumption,'' \emph{Empirical Software Engineering},
  vol.~20, no.~2, pp. 374--409, 2015.

\bibitem{vilajosana2014realistic}
X.~Vilajosana, Q.~Wang, F.~Chraim, T.~Watteyne, T.~Chang, and K.~S. Pister,
  ``\relax {A realistic energy consumption model for TSCH networks},''
  \emph{IEEE Sensors Journal}, vol.~14, no.~2, pp. 482--489, 2014.

\bibitem{wang2006realistic}
Q.~Wang, M.~Hempstead, and W.~Yang, ``A realistic power consumption model for
  wireless sensor network devices,'' in \emph{3rd Annual IEEE Communications
  Society on Sensor and Ad Hoc Communications and Networks (SECON'06)},
  vol.~1.\hskip 1em plus 0.5em minus 0.4em\relax IEEE, 2006, pp. 286--295.

\bibitem{martinez2015power}
B.~Martinez, M.~Monton, I.~Vilajosana, and J.~D. Prades, ``\relax the power of
  models: Modeling power consumption for iot devices,'' \emph{IEEE Sensors
  Journal}, vol.~15, no.~10, pp. 5777--5789, 2015.

\bibitem{kaup2018progress}
F.~Kaup, S.~Hacker, E.~Mentzendorff, C.~Meurisch, and D.~Hausheer, ``The
  progress of the energy-efficiency of single-board computers,'' \emph{Tech.
  Rep. NetSys-TR-2018-01}, 2018.

\bibitem{ARMProfiling}
J.~Nunez-Yanez and G.~Lore, ``\relax{Enabling accurate modeling of power and
  energy consumption in an ARM-based System-on-Chip},'' \emph{Microprocessors
  and Microsystems}, vol.~37, pp. 319--332, 2013.

\bibitem{cramfs}
\BIBentryALTinterwordspacing
\relax{Cramfs: cram a filesystem onto a small ROM}. [Online]. Available:
  \url{https://www.kernel.org/doc/Documentation/filesystems/cramfs.txt}
\BIBentrySTDinterwordspacing

\bibitem{Lim2015EnhancingIS}
G.~Lim, J.~young Hwang, K.~Park, and S.-B. Suh, ``Enhancing init scheme for
  improving bootup time in mobile devices,'' \emph{2015 Eighth International
  Conference on Mobile Computing and Ubiquitous Networking (ICMU)}, pp.
  149--154, 2015.

\bibitem{park2006linux}
C.~Park, K.~Kim, Y.~Jang, and K.~Hyun, ``Linux bootup time reduction for
  digital still camera,'' in \emph{Linux Symposium}, 2006, p. 231.

\bibitem{joe2011bootup}
I.~Joe and S.~C. Lee, ``Bootup time improvement for embedded linux using
  snapshot images created on boot time,'' in \emph{The 2nd International
  Conference on Next Generation Information Technology (ICNIT)}.\hskip 1em plus
  0.5em minus 0.4em\relax IEEE, 2011, pp. 193--196.

\bibitem{baik2010boosting}
K.~Baik, S.~Kim, S.~Woo, and J.~Choi, ``Boosting up embedded linux device:
  experience on linux-based smartphone,'' in \emph{proceedings of the Linux
  Symposium}, 2010, pp. 9--18.

\bibitem{procaccianti2016empirical}
G.~Procaccianti, H.~Fern{\'a}ndez, and P.~Lago, ``Empirical evaluation of two
  best practices for energy-efficient software development,'' \emph{Journal of
  Systems and Software}, vol. 117, pp. 185--198, 2016.

\bibitem{pinto2016comprehensive}
G.~Pinto, K.~Liu, F.~Castor, and Y.~D. Liu, ``A comprehensive study on the
  energy efficiency of java's thread-safe collections,'' in \emph{IEEE
  International Conference on Software Maintenance and Evolution
  (ICSME)}.\hskip 1em plus 0.5em minus 0.4em\relax IEEE, 2016, pp. 20--31.

\bibitem{li2013calculating}
D.~Li, S.~Hao, W.~G. Halfond, and R.~Govindan, ``Calculating source line level
  energy information for android applications,'' in \emph{Proceedings of the
  International Symposium on Software Testing and Analysis}.\hskip 1em plus
  0.5em minus 0.4em\relax ACM, 2013, pp. 78--89.

\bibitem{hao2012estimating}
S.~Hao, D.~Li, W.~G. Halfond, and R.~Govindan, ``Estimating android
  applications' cpu energy usage via bytecode profiling,'' in \emph{Proceedings
  of the First International Workshop on Green and Sustainable Software}.\hskip
  1em plus 0.5em minus 0.4em\relax IEEE, 2012, pp. 1--7.

\end{thebibliography}

\end{document}